\documentclass[aps,preprint,eshowkeys,nofootinbib]{revtex4-1}
\usepackage{verbatim,graphics,graphicx,color,slashed,textcomp,bbm,mathdots,multirow,array}
\usepackage[normalem]{ulem} 
\usepackage[colorlinks=true,linkcolor=red,urlcolor=blue,citecolor=blue]{hyperref}

\usepackage{autobreak}
\usepackage{float}
\usepackage{caption}
\usepackage{subcaption}
\usepackage{ragged2e}  
\graphicspath{{fig/}}

\makeatletter

\newcommand{\Rmnum}[1]{\expandafter\@slowromancap\romannumeral #1@}
\makeatother

\usepackage[colorlinks=true,linkcolor=red,urlcolor=blue,citecolor=blue]{hyperref}

\begin{document}

\title{Prospects for axion dark matter searches at LISA-like interferometers}
\author{Yue-Hui Yao$^{a,c}$, Tingyuan Jiang$^{a,c}$, and Yong Tang$^{a,b,c}$}
\affiliation{\begin{footnotesize}
		${}^a$University of Chinese Academy of Sciences (UCAS), Beijing 100049, China\\
        ${}^b$School of Fundamental Physics and Mathematical Sciences, \\
		Hangzhou Institute for Advanced Study, UCAS, Hangzhou 310024, China \\
		${}^c$International Center for Theoretical Physics Asia-Pacific, Beijing, China
\end{footnotesize}}

\begin{abstract}
Axion or axionlike particles are one of the leading candidates for dark matter. Because of its tiny coupling with photons, axion dark matter in the background can induce distinct phase velocities for light with different parity, an effect known as birefringence.
Here, we propose a modification to the polarization state of the interspacecraft laser link in LISA-like interferometers to make them sensitive to this birefringence effect. We discuss the prospects of using the Sagnac combinations to search for axion dark matter and derive the corresponding sensitivity. With this setup, we show that next-generation laser interferometers in space would have promising sensitivities on the axion-photon coupling with axion mass around $10^{-19} - 10^{-14}~\text{eV}$.
\end{abstract}

\maketitle

\section{Introduction}
Despite its universal gravitational interaction, the nature of dark matter~(DM) remains elusive.
Among various DM candidates, theoretically well-motivated axions~\cite{PhysRevLett.38.1440, PhysRevLett.40.223, PhysRevLett.40.279, Preskill:1982cy, Abbott:1982af,Dine:1982ah} and, more generally, axionlike particles~\cite{Damour_1994a, Damour_1994b, Capozziello_2011} have attracted increasing interest. They can be produced in the early Universe through various mechanisms and may offer potential solutions to small-scale problems in the $\Lambda$CDM~\cite{de_Blok_2009, Boylan_Kolchin_2011, Bullock_2017, Tulin_2018}.

One important feature of axions\footnote{Hereafter, we use ``axions" to collectively refer to both axions and axionlike particles.} is their coupling to gauge bosons, particularly photons~\cite{chadhaday2022axiondarkmatternow}.
This interaction induces photon-axion conversion in the presence of a background magnetic field, with profound implications in astrophysics, such as spectral distortion~\cite{Reynes:2021bpe, Marsh_2017} and rapid stellar cooling~\cite{Giannotti:2017hny, Corsico:2012sh}. It also serves as the working principle for many ground-based experiments, such as CAST~\cite{CAST:2017uph} and ``light shining through a wall"~\cite{PhysRevD.92.092002}.

Another key prediction of the axion-photon coupling is that light with different parity exhibits distinct phase velocities in the axion background, a phenomenon known as birefringence. One result of this difference in phase velocities is the rotation of the polarization plane of linearly polarized light as it propagates through the background. 
Many proposals and experiments leverage this effect, including those based on ground-based cavities~\cite{PhysRevLett.102.202001, PhysRevD.100.023548, PhysRevLett.132.191002, pandey2024resultsaxiondarkmatterbirefringent, PhysRevD.101.095034} and observations of polarized light from celestial bodies~\cite{Ivanov:2018byi, PhysRevD.101.063012, PhysRevD.100.015040, Plascencia:2017kca, PhysRevLett.122.191101, Yuan:2020xui, PhysRevLett.130.121401}.
Powerful constraints and useful sensitivities on the coupling strength have been obtained.

It has been suggested that a gravitational-wave~(GW) interferometer, with appropriate modifications, can also leverage this birefringence effect to search for axions~\cite{PhysRevD.98.035021, PhysRevLett.123.111301, Heinze_2024}, as it can measure light phase with exceptional precision.\footnote{DM may also leave traces in interferometers through other different mechanisms~\cite{Aoki:2016kwl, Vermeulen:2021epa, PhysRevLett.121.061102, LIGOScientific:2021ffg, Guo:2019ker, PhysRevD.107.063015, PhysRevD.108.092010, PhysRevD.108.095054, PhysRevD.108.083007, Kim:2023pkx, Yu:2024enm, PhysRevD.110.095015}.}
Here, we investigate the prospects for searching for axion DM in Laser Interferometer Space Antenna (LISA)-like space-based gravitational-wave interferometers~\cite{amaroseoane2017laserinterferometerspaceantenna, Hu:2017mde, Luo_2016, Crowder:2005nr}. 
The main idea is to replace the linearly polarized light in the interspacecraft laser link with circularly polarized light and utilize the Sagnac combinations to search for axion signals.
With this modification, we show that LISA-like detectors will be sensitive to axion-induced birefringence. 
We estimate the sensitivities of these interferometers and compare them with existing constraints. We find that LISA-like interferometers can be more sensitive to the axion-photon coupling than the existing limits in some mass ranges. In particular, interferometers with smaller noise levels, such as Big Bang Observer~(BBO), could improve the sensitivity by several orders of magnitude across a broad mass range.

The paper is organized as follows. In Sec.~\ref{Axion-induced birefringence}, we present the theoretical framework related to axion-induced birefringence. In Sec.~\ref{Response of interferometer}, we describe the structure of LISA-like detectors and the algorithm to analyze the data stream. We then discuss the single-link response with circularly polarized laser in Sec.~\ref{Circularly polarized light} and linearly polarized laser in Sec.~\ref{Linearly polarized light}. In Sec.~\ref{Modified configuration and Sagnac combination}, we provide a detailed description of the proposed modification and the use of Sagnac combinations to search for axion signals. In Sec.~\ref{Sensitivity}, we derive the sensitivities of the interferometers, compare them with existing bounds, and investigate the factors limiting sensitivity. Finally, we conclude in Sec.~\ref{Conclusion}.

In this paper, we use natural units $c=\hbar=1$ and the metric $g_{\mu\nu}=(+,-,-,-)$.

\section{Axion-induced birefringence} \label{Axion-induced birefringence}
In this section, we present the theoretical framework for the birefringence effect arising from axion-photon interactions as electromagnetic waves propagate through the axion field.

The Lagrangian describing the Chern-Simons interaction between the axion and photon field is given by~\cite{PhysRevD.41.1231, PhysRevD.43.3789, Harari:1992ea}
\begin{equation} \label{axion-photon Lag}
    \mathcal{L} = -\frac{1}{4}F_{\mu\nu}F^{\mu\nu}
    + \frac{1}{2}\partial_{\mu}a\partial^{\mu}a - \frac{1}{2}m^2a^2
    - \frac{g_{a\gamma}}{4}aF_{\mu\nu}\tilde{F}^{\mu\nu},
\end{equation}
where $a$ is the axion field with mass $m$, $F^{\mu\nu}$ and $\tilde{F}^{\mu\nu}=\frac{1}{2}\epsilon^{\mu\nu\rho\sigma}F_{\rho\sigma}$ are the electromagnetic field tensor and its dual tensor, respectively, and $g_{a\gamma}$ is the coupling between axion and photon.
The equations of motion for the fields are derived as
\begin{align}
    \label{Maxwell eq}\partial_{\beta}F^{\beta\alpha} &= -g_{a\gamma}\partial_{\mu}a\tilde{F}^{\mu\alpha},\\
    \left(\partial_{\mu}\partial^{\mu} + m^2\right)a &= -\frac{g_{a\gamma}}{4}F_{\mu\nu}\tilde{F}^{\mu\nu},
\end{align}
which can also be expressed explicitly in terms of the electric and magnetic fields:
\begin{align}
    \label{Gauss law}\nabla\cdot\boldsymbol{E} &= -g_{a\gamma}\nabla a\cdot \boldsymbol{B},\\
    \label{Ampere law}\frac{\partial\boldsymbol{E}}{\partial t} - \nabla\times\boldsymbol{B} &=
    -g_{a\gamma}\left(\frac{\partial a}{\partial t}\cdot\boldsymbol{B} + \nabla a\times\boldsymbol{E}\right),\\
    \label{axion EOM}\left(\partial_{\mu}\partial^{\mu} + m^2\right)a &= g_{a\gamma}\boldsymbol{E}\cdot\boldsymbol{B}.
\end{align}

\begin{figure}[t]
    \centering
    \includegraphics[width=0.7\linewidth]{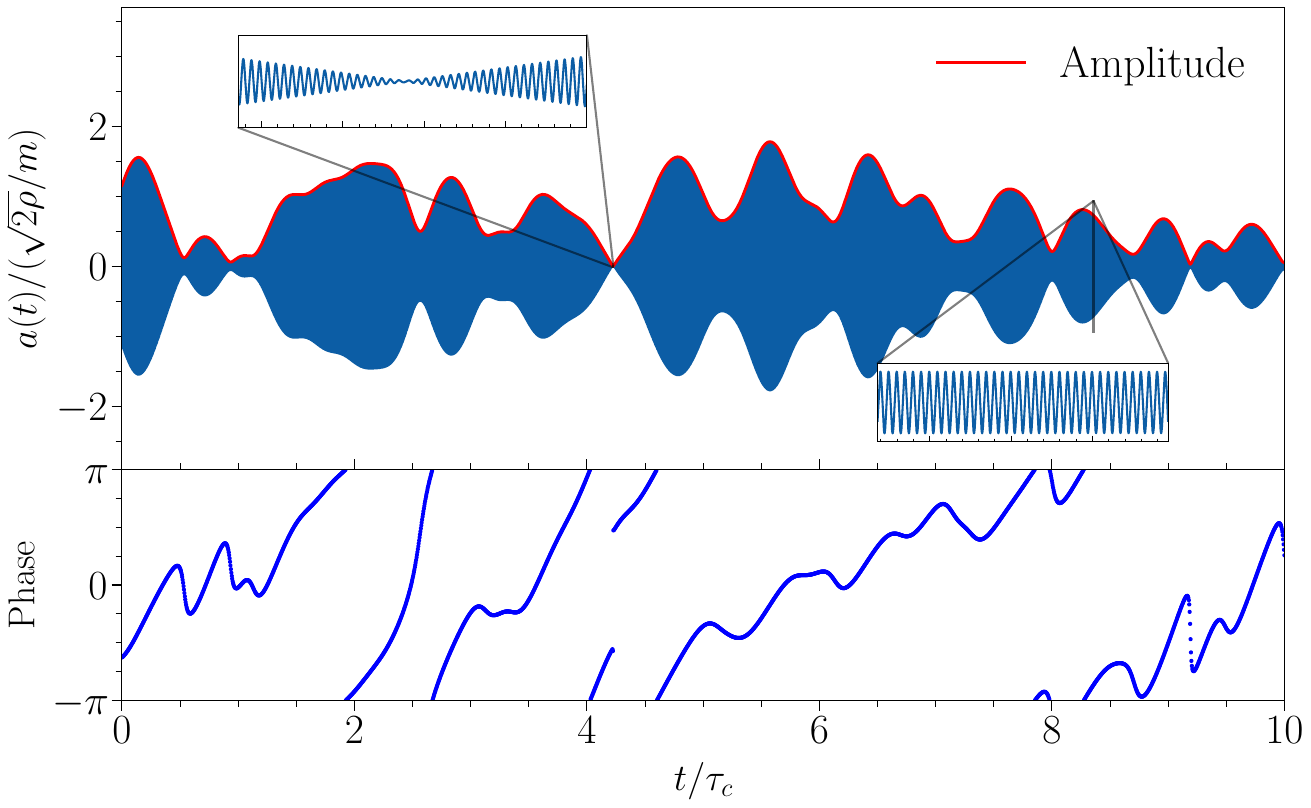}
    \caption{\justifying The evolution of the axion field at a fixed spatial point.
    The amplitude and phase vary stochastically over the coherence time $\tau_c$. The two inset plots show the Compton oscillation.
    }
    \label{fig:field}
\end{figure}

Given that the laser power in the interferometer is only about a few watts and $a\sim\sqrt{2\rho_{\text{DM}}}/m$, where $\rho_{\text{DM}} \approx 0.4~\text{GeV}/\text{cm}^3$ is the local DM energy density, we can neglect the backaction of light on the axion field and set the right-hand side of Eq.~(\ref{axion EOM}) to zero.
The homogeneous Klein-Gordon equation has plane-wave solutions $e^{i\left(\omega_a t - \boldsymbol{k}_a\cdot\boldsymbol{x} + \theta_a\right)}$,
where $\omega_a \simeq m(1+\boldsymbol{v}^2/2)$ and $\boldsymbol{k}_a \simeq m\boldsymbol{v}$, describing the collective motion of axion particles with velocity $\boldsymbol{v}$.
Since DM particles near the solar system exhibit a range of velocities, the axion field can be represented as a superposition of monochromatic plane waves corresponding to axion particles with varying velocities.
The behavior of the axion field can be described by~\cite{Khmelnitsky_2014, Hui:2021tkt}
\begin{equation} \label{axion field}
    a(\boldsymbol{x},t) = a_0(\boldsymbol{x},t)\cos\left(mt+\theta_0\left(\boldsymbol{x}, t\right)\right),
\end{equation}
where the amplitude $a_0(\boldsymbol{x},t)$ and phase $\theta_0(\boldsymbol{x},t)$ vary stochastically on scales set by the coherence time $\tau_c$ and coherence length $\lambda_c$:
\begin{align}
    \label{tc}\tau_c &= \frac{2\pi}{m\sigma^2} \approx 4.13 \times 10^8~\text{s} \left(\frac{10^{-17}~\text{eV}}{m}\right),\\
    \label{lc}\lambda_c &= \frac{2\pi}{m\sigma} \approx 1.24 \times 10^{11}~\text{km} \left(\frac{10^{-17}~\text{eV}}{m}\right),
\end{align}
where $\sigma\simeq10^{-3}$ is the velocity dispersion of local DM.
Figure~\ref{fig:field} illustrates the evolution of the axion field at a fixed spatial point.

We can make further simplifications. Since axion considered here has a coherence length much longer than the interferometer's arm length $\sim\mathcal{O}(10^6~\text{km})$ and a coherence time exceeding the observation duration $\sim\mathcal{O}(1~\text{yr})$, we can neglect the spatial and temporal dependence of the amplitude and phase, and Eq.~(\ref{axion field}) reduces to
\begin{equation}
    a(t) = a_0 \cos(mt+\theta_0) = a_0 e^{i(mt+\theta_0)}.
\end{equation}
Combining Eqs.~(\ref{Gauss law}) and (\ref{Ampere law}) with the homogeneous Maxwell equations\footnote{The homogeneous Maxwell equations are expressed by the Bianchi identities $\partial_{\mu}\tilde{F}^{\mu\nu}=0$ and remain unaltered by the axion-photon coupling.} and neglecting all gradient terms, we obtain the wave equations
\begin{align}
    \label{electric EOM}\left(\frac{\partial^2}{\partial t^2} - \nabla^2\right)\boldsymbol{E} 
    &= g_{a\gamma}\left(\dot{a}\nabla\times\boldsymbol{E}-\ddot{a}\boldsymbol{B}\right),\\
    \left(\frac{\partial^2}{\partial t^2} - \nabla^2\right)\boldsymbol{B}
    &= g_{a\gamma}\dot{a}\nabla\times\boldsymbol{B}.
\end{align}
Since the laser's angular frequency satisfies $\omega\sim\mathcal{O}(\text{eV})\gg m$ and $\ddot{a}\boldsymbol{B}/\dot{a}\nabla\times\boldsymbol{E} \sim m/\omega$, we can neglect the term $\ddot{a}\boldsymbol{B}$ in Eq.~(\ref{electric EOM}), and the wave equations share the same form. 
We exemplify the birefringence effect with the electric field in the following discussion.

The condition $m\ll\omega$ also implies the existence of a sufficiently small spacetime region where $\dot{a}$ can be treated as static, and Eq.~(\ref{electric EOM}) admits a monochromatic plane-wave solution
\begin{equation} \label{plane-wave solution}
    \boldsymbol{E} = \sum_{j=\pm}E_{j} e^{i(\omega t-\boldsymbol{k}\cdot\boldsymbol{x})} \boldsymbol{\epsilon}_{j}(\hat{\boldsymbol{k}}),
\end{equation}
where $E_{\pm}$ are complex amplitudes and $\boldsymbol{\epsilon}_{\pm}$
are the circularly polarized basis satisfying 
$i\hat{\boldsymbol{k}}\times\boldsymbol{\epsilon}_{\pm}(\hat{\boldsymbol{k}}) 
= \pm\boldsymbol{\epsilon}_{\pm}(\hat{\boldsymbol{k}})$.
Substituting Eq.~(\ref{plane-wave solution}) into Eq.~(\ref{electric EOM}), we have the dispersion relations 
\begin{equation} \label{dispersion relation}
    \omega^2 - k^2 = \pm g_{a\gamma}\dot{a}k.
\end{equation}
Consequently, the phase velocities of circularly polarized light are given by
\begin{equation} \label{phase velocity}
    v_{\pm} = \frac{\omega}{k} \simeq 1\pm\frac{g_{a\gamma}\dot{a}}{2\omega}.
\end{equation}
Therefore, left circularly polarized (LCP) and right circularly polarized (RCP) light\footnote{The definitions of LCP and RCP vary among literature. In this work, we refer to $\boldsymbol{\epsilon}_{+}$ as LCP, as the directions of rotation of $\boldsymbol{E}$ and the light's propagation are related by the left-hand rule.} 
are the eigenstates of the axion medium, exhibiting distinct phase velocities as they propagate through it. 
This effect is known as birefringence.

\section{Signal response of interferometer} \label{Response of interferometer}
In this section, we first briefly describe the basic operational principles of LISA-like space-based interferometers. We then derive the single-link response to the axion field, assuming that the light in the link is either linearly polarized~(LP) or circularly polarized~(CP). The main conclusion is that LISA-like interferometers with the current design are insensitive to axion-induced birefringence.

\begin{figure}[t]
    \centering
    \includegraphics[width=0.4\linewidth]{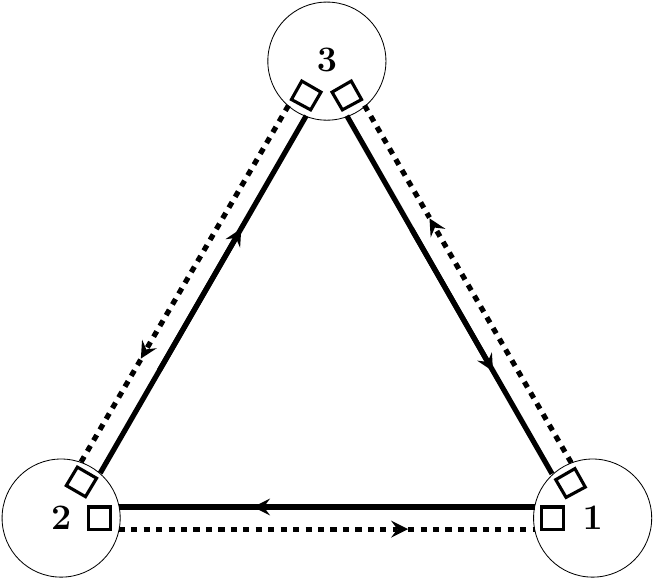}
    \caption{\justifying Schematic representation of LISA. 
    Each spacecraft is equipped with two optical benches. Each spacecraft sends and receives laser beams from the other two, forming three clockwise links and three counterclockwise links.}
    \label{fig:LISA}
\end{figure}

A LISA-like interferometer typically consists of three spacecraft,\footnote{
There are several stages for BBO. In the final stage, there would be four LISA-like constellations distributed along the solar orbit. Here, we focus on the first stage, which involves a single constellation made up of three spacecraft.} forming a quasi-equilateral triangle configuration in space with arm lengths of approximately $\mathcal{O}(10^6\,\text{km})$ for the heliocentric orbit or $\mathcal{O}(10^5\,\text{km})$ for the geocentric orbit.
Each spacecraft contains two optical benches, each hosting a free-falling test mass. Every spacecraft sends and receives laser beams from the other two. The laser from a distant spacecraft interferes with the local laser, forming a beat note whose phase is related to the light travel time between the two spacecraft. Consequently, the three spacecraft result in six laser links with three clockwise and three counterclockwise, as shown in Fig.~\ref{fig:LISA}.

Given the current stability of the laser, the six single-link data streams $\eta_{rs}(t)$ (the nonequal subscripts $r,s=1,2,3$ denote the indices of the spacecrafts) will be overwhelmed by laser noise. Furthermore, due to the orbital motion of the spacecraft, the lengths of the interferometer's arms are time changing and unequal, making the simple Michelson interferometric configuration fail to cancel the laser noise. 
To address these challenges, we employ time-delay interferometry~(TDI)~\cite{Tinto:2020fcc, Armstrong_1999, PhysRevD.62.042002, PhysRevD.65.102002, PhysRevD.72.042003}.
The essence of TDI lies in synthesizing virtual interferometric configurations with nearly equal light path lengths. This is achieved by skillfully time shifting and combining the data streams $\eta(t)_{rs,ij\cdots} \equiv \eta_{rs}(t-L_i-L_j-\cdots)$, for example, Eqs.~(\ref{alpha}) and (\ref{eq:X}). It can be shown that there are numerous laser-noise-free TDI configurations~(also called combinations) with distinct features. 
Among them, two standard categories are the Michelson-like combinations and the Sagnac combinations.
Since the elements of any TDI combination are the single-link data streams, we begin with a detailed study of the single-link response to the axion field.

\subsection{Circularly polarized light} \label{Circularly polarized light}
We start with the case of CP light. Assuming that the light is CP when it leaves the sending spacecraft, 
the time $\Delta T_{\pm}$ taken by the light to reach the receiving spacecraft at time $t$ is given by
\begin{equation}
    L_{rs} = \int^{t}_{t-\Delta T_{\pm}} dt\;v_{\pm}  
    = \Delta T_{\pm} \pm \frac{g_{a\gamma}}{2\omega}\left[a(t)-a(t-\Delta T_{\pm})\right],
\end{equation}
where $L_{rs} \simeq |\boldsymbol{x}_r(t) - \boldsymbol{x}_s(t-L_{rs})|$. Keeping the leading order in $g_{a\gamma}$, we have
\begin{equation} \label{travel time}
    \Delta T_{\pm} \simeq L_{rs} \mp \frac{g_{a\gamma}}{2\omega}\left[a(t)-a(t-L_{rs})\right].
\end{equation}
The relative laser frequency fluctuations induced by axion, or data streams, are given by
\begin{equation} \label{laser fluc}
    \eta_{rs}(t) =
    -\frac{d\left(\Delta T_{\pm}\right)}{dt}
    = \pm \frac{im g_{a\gamma}}{2\omega}\left[a(t)-a(t-L_{rs})\right].
\end{equation}

Note that the axion signal in Eq.~(\ref{laser fluc}) does not depend on the propagation direction of light but only on the difference in the axion field at the emission and reception times. 
This should be contrasted with the scenario where the signal is related to a vector field or the gradient of a scalar field. In that case, the signal involves the scalar product of the field gradient or vector field with the arm's direction vector, i.e., $\hat{\boldsymbol{n}}\cdot\nabla a$ or $\hat{\boldsymbol{n}}\cdot \boldsymbol{A}$, indicating that the signal is direction dependent and will vary annually due to the orbital motion of the constellation~\cite{PhysRevD.103.076018, PhysRevD.104.055037, PhysRevD.105.035029, Amaral:2024tjg}.

\subsection{Linearly polarized light} \label{Linearly polarized light}
Now, we turn to the case where the light is LP when it leaves the spacecraft, which reflects the current design of LISA-like detectors.
\begin{figure}[t]
    \centering
    \includegraphics[width=0.5\linewidth]{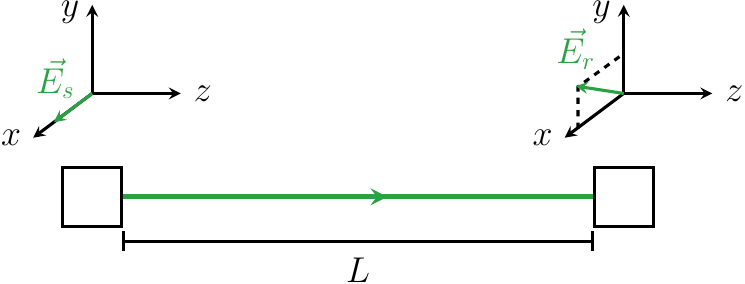}
    \caption{\justifying Schematic representation of the light path for a single link under the current design. Light is horizontally polarized upon leaving the sending spacecraft. As it travels through the axion dark matter background, the polarization angle rotates, developing a vertically polarized component.}
    \label{fig:lightpath linear}
\end{figure}
Without loss of generality, we assume that the light is $x$ polarized (horizontally polarized, HP) in the absence of the axion field.
The schematic plot of the light path is shown in Fig.~\ref{fig:lightpath linear}. The $x$-polarized light is described by the Jones vector~\cite{book}
\begin{equation}
    \boldsymbol{E}_e(t) = \begin{bmatrix}
        1 \\
        0
    \end{bmatrix}e^{i\omega t}
    = \begin{bmatrix}
        1 \\
        i
    \end{bmatrix}\frac{e^{i\omega t}}{2}
    + \begin{bmatrix}
        1 \\
        -i
    \end{bmatrix}\frac{e^{i\omega t}}{2},
\end{equation}
where we expand it into the CP basis. Then, according to Eq.~(\ref{travel time}), the electric field received by the other spacecraft at time $t$ is
\begin{equation}
    \boldsymbol{E}_r(t) = 
    \begin{bmatrix}
        1 \\
        i
    \end{bmatrix} \frac{e^{i\omega(t-\Delta T_+)}}{2}
    + \begin{bmatrix}
        1 \\
        -i
    \end{bmatrix} \frac{e^{i\omega(t-\Delta T_-)}}{2}.
\end{equation}
We can read out the components of $\boldsymbol{E}_r$ in the LP basis directly:
\begin{align}
    \label{E_x response}E_x 
    &= \frac{\cos[\omega(t-\Delta T_+)]+\cos[\omega(t-\Delta T_-)]}{2}\notag \\
    &= \cos\left(\omega\frac{\Delta T_--\Delta T_+}{2}\right) 
    \cos\left[\omega\left(t-\frac{\Delta T_-+\Delta T_+}{2}\right)\right] \\
    &= \cos\left[g_{a\gamma}\frac{a(t)-a(t-L_{rs})}{2}\right]
    \cos\left[\omega\left(t-L_{rs}\right)\right], \notag
\end{align}
and
\begin{align}
    \label{E_y response}E_y 
    &= \frac{-\sin[\omega(t-\Delta T_+)]+\sin[\omega(t-\Delta T_-)]}{2} \notag\\
    &= -\sin\left(\omega\frac{\Delta T_--\Delta T_+}{2}\right) 
    \cos\left[\omega\left(t-\frac{\Delta T_-+\Delta T_+}{2}\right)\right] \\
    &= -\sin\left[g_{a\gamma}\frac{a(t)-a(t-L_{rs})}{2}\right]\cos\left[\omega\left(t-L_{rs}\right)\right].\notag
\end{align}
Equations~(\ref{E_x response}) and (\ref{E_y response}) indicate that, when traveling through the axion field, the polarization angle of the received light varies periodically. This variation in polarization angle produces a vertically polarized~(VP) component, which serves as a target for many experiments~\cite{PhysRevLett.122.191101, PhysRevLett.123.111301, PhysRevLett.132.191002, PhysRevD.100.023548, PhysRevD.101.063012, PhysRevD.101.095034}.
However, the current design of LISA-like interferometers is not sensitive to the evolution of the polarization angle but rather to the phase of the HP component. From Eq.~(\ref{E_x response}), while the axion field causes the amplitude of the HP component to experience periodic modulation at $\mathcal{O}(g_{a\gamma}^2)$, it leaves the phase intact compared to the vacuum case. 
Therefore, within the current design, LISA-like interferometers are insensitive to axion-induced birefringence.

There are two approaches to make the interferometer sensitive to axion-induced birefringence.
One approach involves replacing the current polarization-insensitive design with a polarization-sensitive design~\cite{PhysRevLett.123.111301,PhysRevLett.132.191002}, allowing GW interferometers to search for the variations in the polarization angle. The second approach is to substitute the LP light transmitted between spacecrafts with CP light~\cite{PhysRevD.98.035021}. In the following discussion, we will focus on the latter possibility.

\section{Modified configuration and the Sagnac combinations} \label{Modified configuration and Sagnac combination}

The issue can be addressed by replacing the LP light with CP light in the link. 
This can be achieved by introducing wave plates at suitable positions along the light path.
Here, we propose a modification of the light path and the utilization of the Sagnac combinations in TDI to search for axion-induced signals.

\begin{figure}
  \centering
  \begin{subfigure}[b]{0.45\textwidth}
    \centering
    \includegraphics[width=\linewidth]{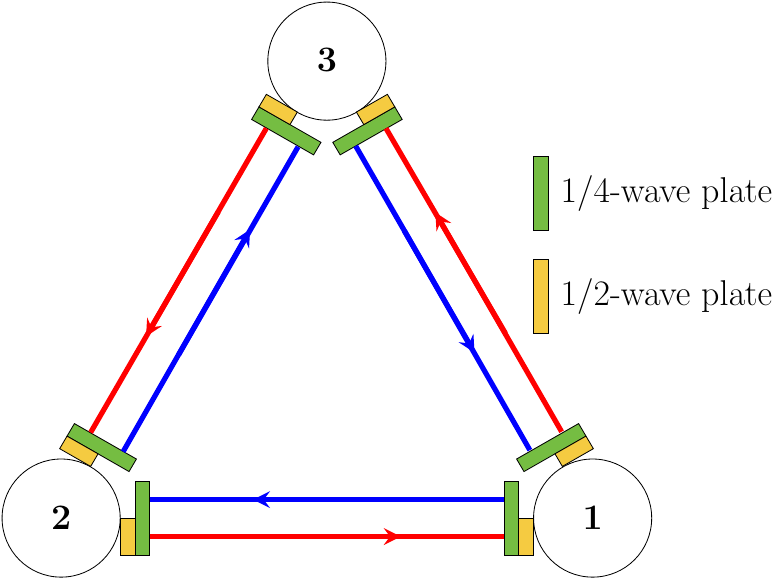}\\
  \end{subfigure}
  \hfill
  \begin{subfigure}[b]{0.45\textwidth}
    \centering
    \includegraphics[width=0.75\linewidth]{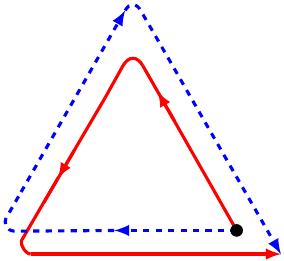}\\
  \end{subfigure}
  \caption{\justifying (a) Schematic representation of the modified light path, with the blue and red lines indicating left circularly polarized light and right circularly polarized light, respectively. Compared to the original design, the linearly polarized light in interspacecraft laser link is replaced with left circularly polarized light (right circularly polarized light) for clockwise (counterclockwise) links. (b) The synthetic light paths of the Sagnac combination $\alpha$.}
  \label{fig:modified LISA}
\end{figure}
The modification is illustrated in Fig.~\ref{fig:modified LISA}. A pair of quarter-wave plate and half-wave plates are added at the ends of each link, which changes the polarization state of the light between horizontally polarized and left circularly polarized (right circularly polarized) when light travels clockwise (counterclockwise).

\begin{figure}
    \centering
    \includegraphics[width=0.5\linewidth]{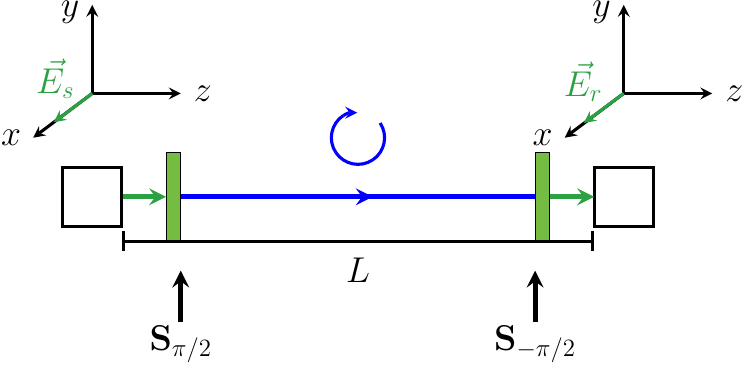}
    \caption{\justifying Schematic representation of the light path for a clockwise link with additional quarter-wave plates. The horizontally polarized light is converted into left circularly polarized light upon leaving the sending spacecraft. Upon reaching the receiving spacecraft, the left circularly polarized is converted back into horizontally polarized light.}
    \label{fig:lightpath waveplate}
\end{figure}

To illustrate the effect of this design, we first consider the signal in a clockwise link.
The light path of the link is shown in Fig.~\ref{fig:lightpath waveplate}.
The Jones matrix for the quarter-wave plate near the sending spacecraft is given by
\begin{equation}
    \mathbf{S}_{\pi/2} 
    = \frac{1}{\sqrt{2}}
    \begin{bmatrix}
        1 & i\\
        i & 1
    \end{bmatrix},
\end{equation}
which converts the HP light into a LCP light. 
The Jones matrix for the quarter-wave plate near the receiving spacecraft is given by
\begin{equation}
    \mathbf{S}_{-\pi/2} 
    = \frac{1}{\sqrt{2}}
    \begin{bmatrix}
        1 & -i \\
        -i & 1
    \end{bmatrix},
\end{equation}
which converts the LCP light back into a HP light.
Since both Jones matrices have $\left|\text{det}[\mathbf{S}]\right|=1$, the inclusion of these additional wave plates does not reduce the laser power.
With this configuration, the electric field received at the end of the link at time $t$ is given by
\begin{equation}
    \boldsymbol{E}_r = 
    \mathbf{S}_{-\pi/2}e^{i\omega(t-\Delta T_+)}\mathbf{S}_{\pi/2}
    \begin{bmatrix}
        1 \\
        0
    \end{bmatrix}
    = \begin{bmatrix}
        1 \\
        0
    \end{bmatrix}e^{i\omega(t-\Delta T_+)},
\end{equation}
where the field is horizontally polarized and has a phase shift $\omega\Delta T_+$.
Therefore, the signal in a clockwise link is given by the positive sign in Eq.~(\ref{laser fluc}).

A counterclockwise link includes an additional half-wave plate, which converts the HP light into a VP light. This VP light is then transformed into RCP light by the quarter-wave plate. At the end of a counterclockwise link, the reverse process occurs, resulting in a HP light with a phase shift $\omega\Delta T_-$, and  
the signal in a counterclockwise link is given by the negative sign in Eq.~(\ref{laser fluc}).

With this design, the synthetic virtual light path of the Sagnac combinations is analogous to that of a ring cavity setup, as shown in the right plot in Fig.~\ref{fig:modified LISA}.
Since the additional phases for the two CP lights differ by a minus sign, and the output of Sagnac combinations represents the phase difference between the clockwise and counterclockwise light paths, this configuration will maximize the interferometer's sensitivity to axions.
In Appendix~\ref{appendix: X comb}, we compare the sensitivity of the Sagnac combinations to
the axion-photon coupling $g_{a\gamma}$ with that of the Michelson-like combinations. 
We find that the Sagnac combinations exhibit better sensitivity.

\section{Sensitivity} \label{Sensitivity}
Now, we derive the sensitivity of the Sagnac combinations to the axion-photon coupling. 

The Sagnac combination $\alpha$ is given by~\cite{Tinto:2020fcc}
\begin{equation} \label{alpha}
    \alpha(t) = \left(\eta_{13}+\eta_{32,2}+\eta_{21,12}\right) - \left(\eta_{12}+\eta_{23,3}+\eta_{31,13}\right),
\end{equation}
where the time-shifted data streams $\eta(t)_{rs,ij} \equiv \eta_{rs}(t-L_i-L_j)$
and $L_i$ denotes the length of the arm opposite spacecraft~$i$.
Substituting Eq.~(\ref{laser fluc}) into Eq.~(\ref{alpha}) and using the approximation of static and equal arm lengths, i.e., $L \equiv L_i = L_j$, we have 
\begin{equation}
    \alpha(t) =  
    \frac{im g_{a\gamma}a_0}{\omega}\left(1-e^{-i3mL}\right)e^{i(mt+\theta_0)}.
\end{equation}
The one-sided power spectral density~(PSD) of the signal is given by
\begin{equation} \label{signal PSD}
    P_{s}(f) = 2\frac{|\tilde{\alpha}(f)|^2}{T} = 
    \begin{cases}
         \frac{4\rho_{\text{DM}}g^2_{a\gamma}}{\omega^2} \sin^2\left(3\pi fL\right) T,
        \quad &f = f_c;\\
        0, \quad &f \neq f_c,
    \end{cases}
\end{equation}
where $f_c=m/2\pi$ is the axion's Compton frequency, $T$ is the observation duration, and we replace $a^2_0$ with its ensemble average, $\left\langle a^2_0\right\rangle=2\rho_{\text{DM}}/m^2$. 
It is worth noting that Eq.~(\ref{signal PSD}) is valid only for $\tau_c > T$. 
Taking $T = 1~\text{yr}$, this corresponds to axion masses less than a few $10^{-16}~\text{eV}$.
For an axion with a heavier mass and $\tau_c \leq T$, the axion field can no longer be treated as coherent. Consequently, the signal appears as a continuous spectrum centered around $f_c$ with a width of $1/\tau_c$, instead of a sharp spike as in Eq.~(\ref{signal PSD}).
In this case, the magnitude of the signal PSD at $f_c$ is approximately given by~\cite{PhysRevX.4.021030}
\begin{equation} \label{signal PSD incoherent}
    P_{s}(f_c) \simeq \frac{4\rho_{\text{DM}}g^2_{a\gamma}}{\omega^2} \sin^2\left(3\pi f_cL\right) \tau_c ,
\end{equation}
which no longer grows with time.

For $\tau_c > T$, the signal-to-noise ratio~(SNR) is defined as
\begin{equation} \label{snr co}
    \text{SNR} = \frac{P_{s}(f_c)}{P_n(f_c)},
\end{equation}
where $P_{s}(f)$ is given by Eq.~(\ref{signal PSD}) and $P_n(f)$ is the one-sided PSD of the noise in the Sagnac combinations~\cite{PhysRevD.108.083007}
\begin{equation} \label{noise PSD}
    P_n(f) = 8\left[ 2\sin^{2}\left(\pi fL\right) + \sin^{2} \left(3\pi fL\right) \right]  S_{\text{acc}}(f)+ 6S_{\text{oms}}(f) ,
\end{equation}
where $S_{\text{oms}}$ and $S_{\text{acc}}$ denote optical metrology system noise and test mass acceleration noise, respectively. For LISA, Taiji, and TianQin , they are given by~\cite{babak2021lisasensitivitysnrcalculations}
\begin{align}
    \label{oms} S_{\text{oms}}\left( f\right) &= \left(\frac{2\pi f}{c}s_{\text{oms}}\right)^{2} \left[1 + \left(\frac{2 \times 10^{-3}~\textrm{Hz}}{f} \right)^{4}\right]
    \;\frac{1}{\textrm{Hz}}, \\
    \label{acc} S_{\text{acc}}\left( f\right) &= \left( \frac{s_{\text{acc}}}{2\pi fc} \right)^{2} \left[ 1+\left(\frac{0.4 \times 10^{-3}~\textrm{Hz}}{f} \right)^{2} \right] \; \left[ 1+ \left(\frac{f}{8 \times 10^{-3}~\textrm{Hz}} \right)^{4} \right]
    \;\frac{1}{\textrm{Hz}},
\end{align}
while for BBO, the frequency-dependent factors in the square brackets are neglected~\cite{Corbin:2005ny}.
For $\tau_c \leq T$, the SNR is defined as~\cite{PhysRevX.4.021030}
\begin{equation} \label{snr inco}
    \text{SNR} = \frac{P_{s}(f_c)}{P_n(f_c)/\sqrt{T/\tau_c}} ,
\end{equation}
with $P_{s}(f_c)$ given by Eq.~(\ref{signal PSD incoherent}).
The factor $\sqrt{T/\tau_c}$ arises because 
we can ``average" Eq.~(\ref{snr co}) over additional frequency bins around $f_c$ when the signal spectrum is resolved.

\begin{figure}[t]
    \centering
    \includegraphics[width=0.9\linewidth]{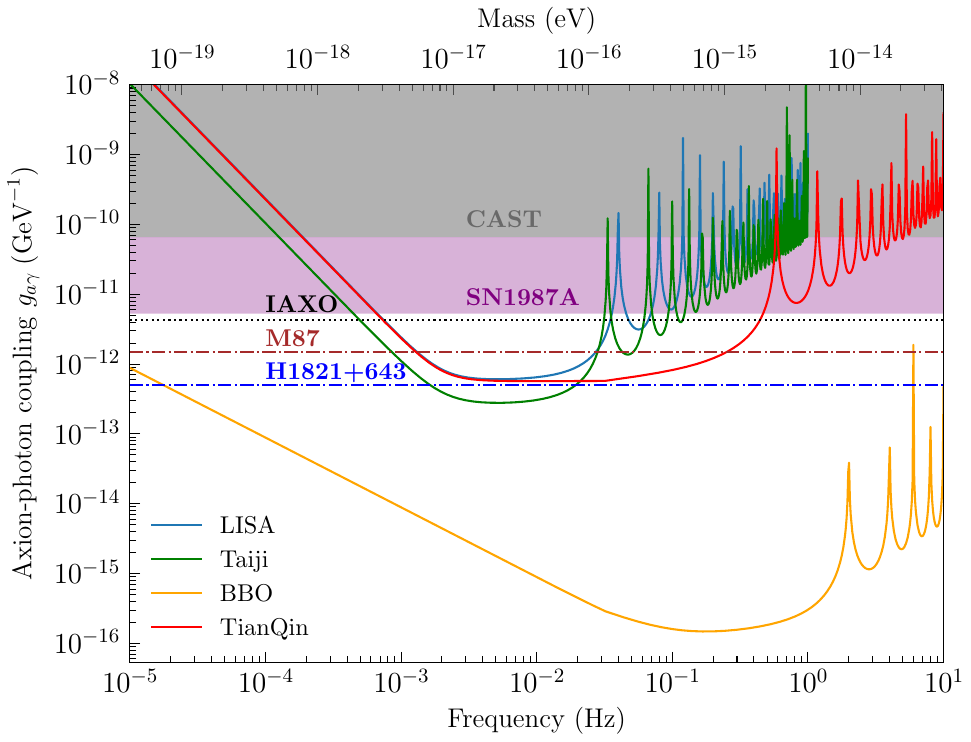}
    \caption{\justifying Sensitivities of several LISA-like interferometers to the axion-photon coupling $g_{a\gamma}$. The bounds from CAST~\cite{CAST:2017uph} and SN1987A~\cite{Payez_2015} are indicated by the gray and purple shaded areas, respectively. The constraints from astrophysical observations of the M87 galaxy~\cite{Marsh_2017} and quasar H1821+643~\cite{Reynes:2021bpe} are shown with dot-dashed lines. The sensitivity of future axion experiment IAXO~\cite{IAXO:2019mpb} is also indicated (black dotted line).}
    \label{fig:sensitivity}
\end{figure}

We define the sensitivity as the coupling strength that results in $\text{SNR}=1$
within a one-year observation period.\footnote{The stochastic nature of the field will generally degrade sensitivity~\cite{Centers:2019dyn, PhysRevD.110.095015}. However, the exact degree of degradation depends on the specific relationship between the signal and the field, as well as the statistical framework used. In this sense, our $\text{SNR}=1$ sensitivity serves as an estimate of the interferometer performance.} 
In Fig.~\ref{fig:sensitivity}, we plot the sensitivities of the four interferometers to the axion-photon coupling $g_{a\gamma}$ and compare them with the existing constraints from the ground-based CAST experiment~\cite{CAST:2017uph} and astrophysical observations~\cite{Payez_2015, Marsh_2017, Reynes:2021bpe}. The parameters adopted for LISA, Taiji, TianQin, and BBO are summarized in Table~\ref{tab:noise para}. Other similar projects DECIGO~\cite{Kawamura_2006}, LISAmax~\cite{Martens_2023}, and ASTROD-GW~\cite{ASTROD-GW} are not listed here. 

\begin{table}
    \centering
    \begin{tabular}{p{4cm}p{2cm}p{2cm}p{2cm}p{2cm}}
    \hline\hline
        ~ & LISA & Taiji & TianQin & BBO \\ \hline
        $\text{Arm length}~L\left(10^9~\text{m}\right)$ & 2.5 & 3 & 0.17 & 0.05 \\ 
        $s_{\text{oms}}\left(10^{-12}~\text{m}\right)$ & 15 & 8 & 1 & $1.4 \times 10^{-5}$ \\ 
        $s_{\text{acc}}\left(10^{-15}~\text{m}\cdot\text{s}^{-2}\right)$ & 3 & 3 & 1 & $3\times 10^{-2}$ \\ \hline\hline
    \end{tabular}
    \caption{\justifying The parameters of several planned space-based interferometers. 
    The laser wavelength is $1064~\text{nm}$ for LISA, Taiji, and TianQin and $355~\text{nm}$ for BBO~\cite{Harry:2006fi}.
    The symbols $s_{\text{oms}}$ and $s_{\text{acc}}$ denote the parameters of optical metrology system noise and test mass acceleration noise, respectively.}
    \label{tab:noise para}
\end{table}

As shown in Fig.~\ref{fig:sensitivity}, in their most sensitive mass range those interferometers will improve on the existing constraints from CAST~\cite{CAST:2017uph} and SN1987A~\cite{Payez_2015}. LISA, Taiji, and TianQin could achieve sensitivities of $ g_{a\gamma}\sim 3\times10^{-13}~\text{GeV}^{-1}$ for axions with mass around $10^{-17}~\text{eV}$.
Because of its better noise performance, BBO could achieve $g_{a\gamma}\sim 10^{-16}~\text{GeV}^{-1}$ across a broad mass range, improving the constraints by 5 orders of magnitude.\footnote{In the final stage of BBO, one can correlate data from the four independent constellations. Then the overall sensitivity would be improved by a factor of two compared to the sensitivity presented here~\cite{PhysRevD.110.095015}.}
We also present the constraints derived from the spectra of active galactic nuclei~\cite{Marsh_2017,Reynes:2021bpe}. While these constraints are comparable to the sensitivities of LISA, Taiji, and TianQin,
they rely on theoretical assumptions about galactic magnetic models.

\begin{figure}
    \centering
    \includegraphics[width=1\linewidth]{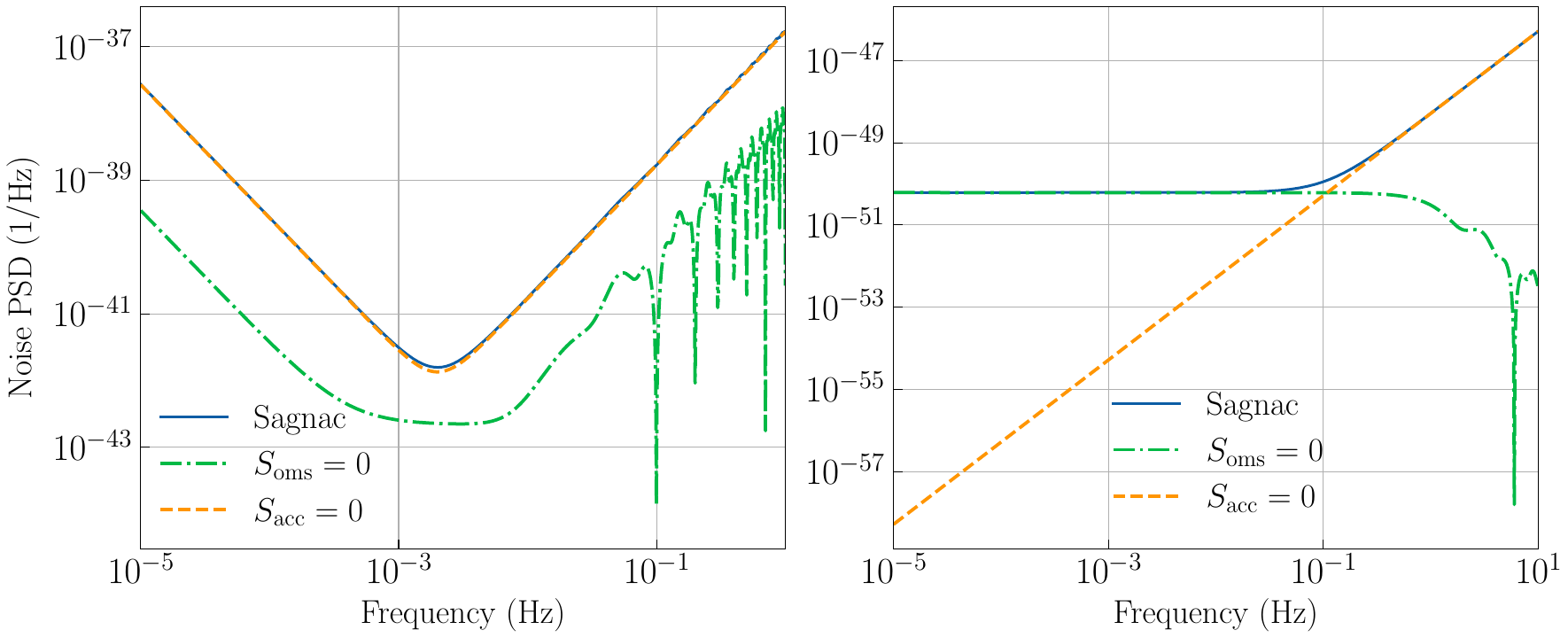}
    \caption{\justifying The noise PSDs in the Sagnac combinations, Taiji (left) and BBO (right). The blue solid line is the full Sagnac noise PSD, while the green dot-dashed line and the orange dashed line show the cases where $S_{\text{oms}}=0$ and $S_{\text{acc}}=0$, respectively.}
    \label{fig:noise}
\end{figure}

Note that, for LISA, Taiji, and TianQin,
the symmetric structure of the Sagnac combinations strongly suppresses acceleration noise, rendering optical metrology noise dominant across the entire frequency band, as shown in Fig.~\ref{fig:noise}. 
Therefore, improving the performance of optical metrology noise is crucial to these interferometers' sensitivity to axions.

Neglecting acceleration noise and working under the low-frequency limit, the sensitivities of LISA, Taiji, and TianQin can be approximated by
\begin{equation} \label{alph seni lw}
    g_{a\gamma} 
    \simeq 
    \omega\left(\frac{S_{\text{oms}}}{6\rho_{\text{DM}}T}\right)^{1/2}\frac{1}{\pi fL}.
\end{equation}
Therefore, an interferometer with a longer arm length and lower optical metrology noise exhibits better sensitivity.
This also explains why the sensitivities of LISA and TianQin are comparable: while LISA benefits from a longer arm length, TianQin achieves a lower optical metrology noise. However, the above discussion does not fully apply to BBO. 
Because, while acceleration noise is suppressed in the Sagnac combination, it still dominates at low frequencies due to BBO’s much improved optical metrology noise, as shown in Fig.~\ref{fig:noise}.

It is worth noting that, while our setup should theoretically preserve laser power, practical wave plates are imperfect, absorbing and reflecting some light, which reduces the laser power.
This will increase optical metrology noise, as it is primarily dominated by shot noise, whose strength is inversely proportional to the number of photons received by the detector. 
Consequently, if the laser power is reduced to $\lambda$ of its original value due to the wave plates,  
the optical metrology noise increases to $S'_{\text{oms}} = S_{\text{oms}}/\lambda$,
and the sensitivity degrades to $g_{a\gamma}/\sqrt{\lambda}$, according to Eq.~(\ref{alph seni lw}).
Since high-quality wave plates can achieve $\lambda \approx 99\%$, the laser power loss is expected to be well controlled, leading to only moderate degradation in sensitivity.

\section{Conclusion} \label{Conclusion}
In this work, we propose to adjust LISA-like gravitational-wave laser interferometers in space to make them also sensitive to axion dark matter. This adjustment changes the polarization of the interspacecraft laser from linearly polarized to circularly polarized, leveraging axion-induced birefringence. We suggest using the laser-noise-free Sagnac combinations to search for axion signals and derive the corresponding sensitivity to the axion-photon coupling $g_{a\gamma}$. We show that LISA-like interferometers could probe previously unexplored regions in parameter space. For LISA, Taiji, and TianQin, these interferometers could improve the sensitivity from CAST and SN1987A by an order of the magnitude in the mass range $10^{-17} - 10^{-15}~\text{eV}$ with a one-year observation. 
Next-generation detector BBO could improve by 5 orders of magnitude with a mass around $10^{-15}~\text{eV}$ due to its much better noise control. Our results indicate that, with minor modifications, laser interferometers in space could significantly advance axion dark matter search, providing other insight for extending the scientific goals of future gravitational-wave detection missions.

\section*{acknowledgement}
The authors are grateful to Yan Cao and Yu-Feng Zhou for insightful discussions. This work is supported by the National Key Research and Development Program of China (Grant No.~2021YFC2201901), and the Fundamental Research Funds for the Central Universities. 

\emph{Note added}: While we were finalizing the manuscript, a preprint~\cite{wolf2024} discussing similar topics appeared on arXiv.

\appendix
\section{The sensitivity of the Michelson-like combinations} \label{appendix: X comb}
\begin{figure}
    \centering
    \includegraphics[width=0.8\linewidth]{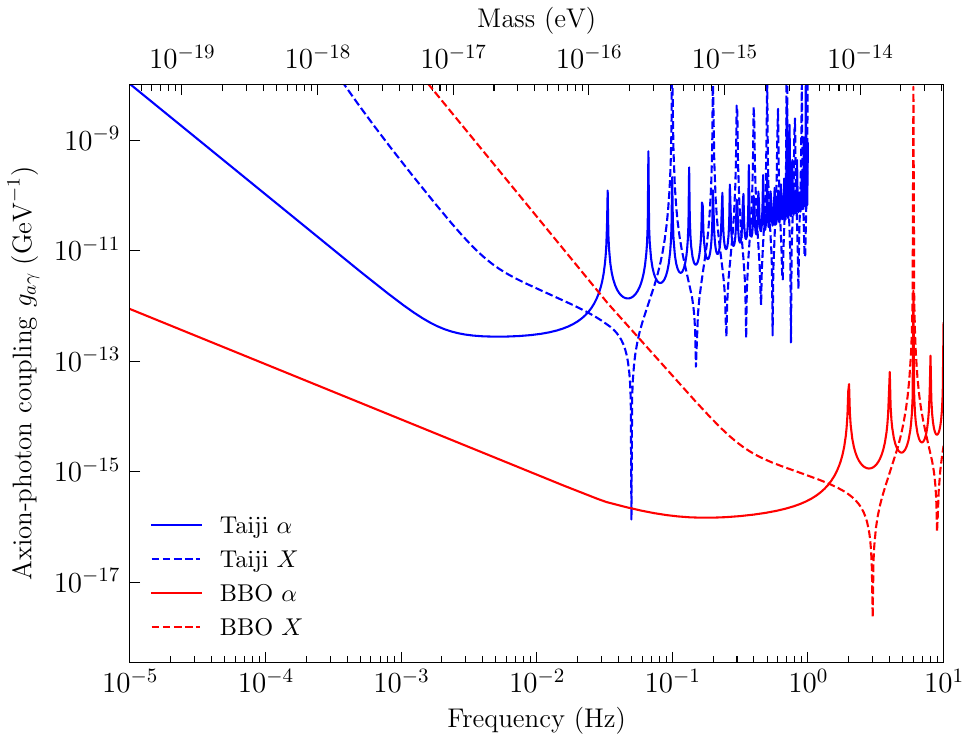}
    \caption{\justifying Sensitivities of Taiji~(blue lines) and BBO~(red lines) using the Sagnac combination~(solid lines) and the Michelson-like combination~(dashed lines), respectively.}
    \label{fig:sensitivity_Sa_X}
\end{figure}

The Michelson-like combination $X$ is given by~\cite{Tinto:2020fcc}
\begin{equation}\label{eq:X}
    X(t) = \left(\eta_{13} + \eta_{31,2} + \eta_{12,22} + \eta_{21,322}\right)
    - \left(\eta_{12} + \eta_{21,3} + \eta_{13,33} + \eta_{31,233}\right) . 
\end{equation}
Substituting Eq.~(\ref{laser fluc}) into the above expression and assuming equal arm lengths, we have
\begin{equation}
    X(t) = 
    \frac{im g_{a\gamma}a_0}{\omega}
    \left(1 - 3e^{-imL} + 3e^{-i2mL} - e^{-i3mL}\right)e^{i(mt+\theta_0)}. 
\end{equation}
Then, for $\tau_c > T$, the one-sided PSD of the signal in the $X$ combination is given by
\begin{equation} \label{X signal PSD}
    P^X_{s}(f) = 2\frac{|\tilde{X}(f)|^2}{T} = 
    \begin{cases}
         \frac{64\rho_{\text{DM}}g^2_{a\gamma}}{\omega^2}  \sin^6\left(\pi fL\right) T,
        \quad &f = f_c;\\
        0, \quad &f \neq f_c .
    \end{cases}
\end{equation}
For $\tau_c \leq T$, the magnitude of the signal PSD at $f_c$ is given by
replacing $T$ with $\tau_c$ in Eq.~(\ref{X signal PSD}).
The SNR is still defined by Eqs.~(\ref{snr co}) and (\ref{snr inco}) for the cases $\tau_c > T$ and $\tau_c \leq T$, respectively, but with the noise PSD of the Michelson-like combination~\cite{PhysRevD.108.083007}  
\begin{equation}~\label{X noise PSD}
    P^X_n(f) = 16\sin^{2}(2\pi fL) 
    \left\{\left[ 3+\cos(4\pi fL) \right]S_{\text{acc}}(f) + S_{\text{oms}}(f)\right\} .
\end{equation}
In Fig.~\ref{fig:sensitivity_Sa_X}, we compare the sensitivities of the $\alpha$ combination and $X$ combination, where the parameters of Taiji and BBO are adopted.
The deep valleys in the sensitivity curve of the $X$ combination 
appear at frequencies $f = n/2L\;(n \in 1,2,\cdots)$, where the noise PSD in Eq.~(\ref{X noise PSD}) vanishes while the signal PSD in Eq.~(\ref{X signal PSD}) remains finite.

Clearly, while their performance is comparable in the high-frequency region, 
the Sagnac combination has better sensitivity across a wider range of frequencies.
This can be understood as follows.
In the low-frequency limit ($2\pi fL < 1$), the sensitivity of the $X$ combination can be approximated by
\begin{equation} \label{X sensi lw}
    g^X_{a\gamma} 
    \simeq 
    2\omega \left(\frac{S_{\text{acc}}}{\rho_{\text{DM}}T}\right)^{1/2}
    \left(\frac{1}{\pi fL}\right)^2 .
\end{equation}
Using Eqs.~(\ref{alph seni lw}) and (\ref{X sensi lw}), the ratio between the sensitivities 
of the $\alpha$ combination and $X$ combination is about
\begin{equation}
    \frac{g^X_{a\gamma}}{g^{\alpha}_{a\gamma}}
    \sim \left(\frac{S_{\text{acc}}}{S_{\text{oms}}}\right)^{1/2}
    \frac{1}{\pi fL} ,
\end{equation}
which is greater than one since the acceleration noise dominates
over the optical noise in the low-frequency regime.

\bibliography{ref}

\begin{thebibliography}{77}%
\makeatletter
\providecommand \@ifxundefined [1]{%
 \@ifx{#1\undefined}
}%
\providecommand \@ifnum [1]{%
 \ifnum #1\expandafter \@firstoftwo
 \else \expandafter \@secondoftwo
 \fi
}%
\providecommand \@ifx [1]{%
 \ifx #1\expandafter \@firstoftwo
 \else \expandafter \@secondoftwo
 \fi
}%
\providecommand \natexlab [1]{#1}%
\providecommand \enquote  [1]{``#1''}%
\providecommand \bibnamefont  [1]{#1}%
\providecommand \bibfnamefont [1]{#1}%
\providecommand \citenamefont [1]{#1}%
\providecommand \href@noop [0]{\@secondoftwo}%
\providecommand \href [0]{\begingroup \@sanitize@url \@href}%
\providecommand \@href[1]{\@@startlink{#1}\@@href}%
\providecommand \@@href[1]{\endgroup#1\@@endlink}%
\providecommand \@sanitize@url [0]{\catcode `\\12\catcode `\$12\catcode
  `\&12\catcode `\#12\catcode `\^12\catcode `\_12\catcode `\%12\relax}%
\providecommand \@@startlink[1]{}%
\providecommand \@@endlink[0]{}%
\providecommand \url  [0]{\begingroup\@sanitize@url \@url }%
\providecommand \@url [1]{\endgroup\@href {#1}{\urlprefix }}%
\providecommand \urlprefix  [0]{URL }%
\providecommand \Eprint [0]{\href }%
\providecommand \doibase [0]{http://dx.doi.org/}%
\providecommand \selectlanguage [0]{\@gobble}%
\providecommand \bibinfo  [0]{\@secondoftwo}%
\providecommand \bibfield  [0]{\@secondoftwo}%
\providecommand \translation [1]{[#1]}%
\providecommand \BibitemOpen [0]{}%
\providecommand \bibitemStop [0]{}%
\providecommand \bibitemNoStop [0]{.\EOS\space}%
\providecommand \EOS [0]{\spacefactor3000\relax}%
\providecommand \BibitemShut  [1]{\csname bibitem#1\endcsname}%
\let\auto@bib@innerbib\@empty
\bibitem [{\citenamefont {Peccei}\ and\ \citenamefont
  {Quinn}(1977)}]{PhysRevLett.38.1440}%
  \BibitemOpen
  \bibfield  {author} {\bibinfo {author} {\bibfnamefont {R.~D.}\ \bibnamefont
  {Peccei}}\ and\ \bibinfo {author} {\bibfnamefont {H.~R.}\ \bibnamefont
  {Quinn}},\ }\href {\doibase 10.1103/PhysRevLett.38.1440} {\bibfield
  {journal} {\bibinfo  {journal} {Phys. Rev. Lett.}\ }\textbf {\bibinfo
  {volume} {38}},\ \bibinfo {pages} {1440} (\bibinfo {year}
  {1977})}\BibitemShut {NoStop}%
\bibitem [{\citenamefont {Weinberg}(1978)}]{PhysRevLett.40.223}%
  \BibitemOpen
  \bibfield  {author} {\bibinfo {author} {\bibfnamefont {S.}~\bibnamefont
  {Weinberg}},\ }\href {\doibase 10.1103/PhysRevLett.40.223} {\bibfield
  {journal} {\bibinfo  {journal} {Phys. Rev. Lett.}\ }\textbf {\bibinfo
  {volume} {40}},\ \bibinfo {pages} {223} (\bibinfo {year} {1978})}\BibitemShut
  {NoStop}%
\bibitem [{\citenamefont {Wilczek}(1978)}]{PhysRevLett.40.279}%
  \BibitemOpen
  \bibfield  {author} {\bibinfo {author} {\bibfnamefont {F.}~\bibnamefont
  {Wilczek}},\ }\href {\doibase 10.1103/PhysRevLett.40.279} {\bibfield
  {journal} {\bibinfo  {journal} {Phys. Rev. Lett.}\ }\textbf {\bibinfo
  {volume} {40}},\ \bibinfo {pages} {279} (\bibinfo {year} {1978})}\BibitemShut
  {NoStop}%
\bibitem [{\citenamefont {Preskill}\ \emph {et~al.}(1983)\citenamefont
  {Preskill}, \citenamefont {Wise},\ and\ \citenamefont
  {Wilczek}}]{Preskill:1982cy}%
  \BibitemOpen
  \bibfield  {author} {\bibinfo {author} {\bibfnamefont {J.}~\bibnamefont
  {Preskill}}, \bibinfo {author} {\bibfnamefont {M.~B.}\ \bibnamefont {Wise}},
  \ and\ \bibinfo {author} {\bibfnamefont {F.}~\bibnamefont {Wilczek}},\ }\href
  {\doibase 10.1016/0370-2693(83)90637-8} {\bibfield  {journal} {\bibinfo
  {journal} {Phys. Lett. B}\ }\textbf {\bibinfo {volume} {120}},\ \bibinfo
  {pages} {127} (\bibinfo {year} {1983})}\BibitemShut {NoStop}%
\bibitem [{\citenamefont {Abbott}\ and\ \citenamefont
  {Sikivie}(1983)}]{Abbott:1982af}%
  \BibitemOpen
  \bibfield  {author} {\bibinfo {author} {\bibfnamefont {L.~F.}\ \bibnamefont
  {Abbott}}\ and\ \bibinfo {author} {\bibfnamefont {P.}~\bibnamefont
  {Sikivie}},\ }\href {\doibase 10.1016/0370-2693(83)90638-X} {\bibfield
  {journal} {\bibinfo  {journal} {Phys. Lett. B}\ }\textbf {\bibinfo {volume}
  {120}},\ \bibinfo {pages} {133} (\bibinfo {year} {1983})}\BibitemShut
  {NoStop}%
\bibitem [{\citenamefont {Dine}\ and\ \citenamefont
  {Fischler}(1983)}]{Dine:1982ah}%
  \BibitemOpen
  \bibfield  {author} {\bibinfo {author} {\bibfnamefont {M.}~\bibnamefont
  {Dine}}\ and\ \bibinfo {author} {\bibfnamefont {W.}~\bibnamefont
  {Fischler}},\ }\href {\doibase 10.1016/0370-2693(83)90639-1} {\bibfield
  {journal} {\bibinfo  {journal} {Phys. Lett. B}\ }\textbf {\bibinfo {volume}
  {120}},\ \bibinfo {pages} {137} (\bibinfo {year} {1983})}\BibitemShut
  {NoStop}%
\bibitem [{\citenamefont {Damour}\ and\ \citenamefont
  {Polyakov}(1994{\natexlab{a}})}]{Damour_1994a}%
  \BibitemOpen
  \bibfield  {author} {\bibinfo {author} {\bibfnamefont {T.}~\bibnamefont
  {Damour}}\ and\ \bibinfo {author} {\bibfnamefont {A.~M.}\ \bibnamefont
  {Polyakov}},\ }\href {\doibase 10.1007/bf02106709} {\bibfield  {journal}
  {\bibinfo  {journal} {General Relativity and Gravitation}\ }\textbf {\bibinfo
  {volume} {26}},\ \bibinfo {pages} {1171–1176} (\bibinfo {year}
  {1994}{\natexlab{a}})}\BibitemShut {NoStop}%
\bibitem [{\citenamefont {Damour}\ and\ \citenamefont
  {Polyakov}(1994{\natexlab{b}})}]{Damour_1994b}%
  \BibitemOpen
  \bibfield  {author} {\bibinfo {author} {\bibfnamefont {T.}~\bibnamefont
  {Damour}}\ and\ \bibinfo {author} {\bibfnamefont {A.}~\bibnamefont
  {Polyakov}},\ }\href {\doibase 10.1016/0550-3213(94)90143-0} {\bibfield
  {journal} {\bibinfo  {journal} {Nuclear Physics B}\ }\textbf {\bibinfo
  {volume} {423}},\ \bibinfo {pages} {532–558} (\bibinfo {year}
  {1994}{\natexlab{b}})}\BibitemShut {NoStop}%
\bibitem [{\citenamefont {Capozziello}\ and\ \citenamefont
  {De~Laurentis}(2011)}]{Capozziello_2011}%
  \BibitemOpen
  \bibfield  {author} {\bibinfo {author} {\bibfnamefont {S.}~\bibnamefont
  {Capozziello}}\ and\ \bibinfo {author} {\bibfnamefont {M.}~\bibnamefont
  {De~Laurentis}},\ }\href {\doibase 10.1016/j.physrep.2011.09.003} {\bibfield
  {journal} {\bibinfo  {journal} {Physics Reports}\ }\textbf {\bibinfo {volume}
  {509}},\ \bibinfo {pages} {167–321} (\bibinfo {year} {2011})}\BibitemShut
  {NoStop}%
\bibitem [{\citenamefont {de~Blok}(2009)}]{de_Blok_2009}%
  \BibitemOpen
  \bibfield  {author} {\bibinfo {author} {\bibfnamefont {W.~J.~G.}\
  \bibnamefont {de~Blok}},\ }\href {\doibase 10.1155/2010/789293} {\bibfield
  {journal} {\bibinfo  {journal} {Advances in Astronomy}\ }\textbf {\bibinfo
  {volume} {2010}} (\bibinfo {year} {2009}),\ 10.1155/2010/789293}\BibitemShut
  {NoStop}%
\bibitem [{\citenamefont {Boylan-Kolchin}\ \emph {et~al.}(2011)\citenamefont
  {Boylan-Kolchin}, \citenamefont {Bullock},\ and\ \citenamefont
  {Kaplinghat}}]{Boylan_Kolchin_2011}%
  \BibitemOpen
  \bibfield  {author} {\bibinfo {author} {\bibfnamefont {M.}~\bibnamefont
  {Boylan-Kolchin}}, \bibinfo {author} {\bibfnamefont {J.~S.}\ \bibnamefont
  {Bullock}}, \ and\ \bibinfo {author} {\bibfnamefont {M.}~\bibnamefont
  {Kaplinghat}},\ }\href {\doibase 10.1111/j.1745-3933.2011.01074.x} {\bibfield
   {journal} {\bibinfo  {journal} {Monthly Notices of the Royal Astronomical
  Society: Letters}\ }\textbf {\bibinfo {volume} {415}},\ \bibinfo {pages}
  {L40–L44} (\bibinfo {year} {2011})}\BibitemShut {NoStop}%
\bibitem [{\citenamefont {Bullock}\ and\ \citenamefont
  {Boylan-Kolchin}(2017)}]{Bullock_2017}%
  \BibitemOpen
  \bibfield  {author} {\bibinfo {author} {\bibfnamefont {J.~S.}\ \bibnamefont
  {Bullock}}\ and\ \bibinfo {author} {\bibfnamefont {M.}~\bibnamefont
  {Boylan-Kolchin}},\ }\href {\doibase 10.1146/annurev-astro-091916-055313}
  {\bibfield  {journal} {\bibinfo  {journal} {Annual Review of Astronomy and
  Astrophysics}\ }\textbf {\bibinfo {volume} {55}},\ \bibinfo {pages}
  {343–387} (\bibinfo {year} {2017})}\BibitemShut {NoStop}%
\bibitem [{\citenamefont {Tulin}\ and\ \citenamefont {Yu}(2018)}]{Tulin_2018}%
  \BibitemOpen
  \bibfield  {author} {\bibinfo {author} {\bibfnamefont {S.}~\bibnamefont
  {Tulin}}\ and\ \bibinfo {author} {\bibfnamefont {H.-B.}\ \bibnamefont {Yu}},\
  }\href {\doibase 10.1016/j.physrep.2017.11.004} {\bibfield  {journal}
  {\bibinfo  {journal} {Physics Reports}\ }\textbf {\bibinfo {volume} {730}},\
  \bibinfo {pages} {1–57} (\bibinfo {year} {2018})}\BibitemShut {NoStop}%
\bibitem [{\citenamefont {Chadha-Day}\ \emph {et~al.}(2022)\citenamefont
  {Chadha-Day}, \citenamefont {Ellis},\ and\ \citenamefont
  {Marsh}}]{chadhaday2022axiondarkmatternow}%
  \BibitemOpen
  \bibfield  {author} {\bibinfo {author} {\bibfnamefont {F.}~\bibnamefont
  {Chadha-Day}}, \bibinfo {author} {\bibfnamefont {J.}~\bibnamefont {Ellis}}, \
  and\ \bibinfo {author} {\bibfnamefont {D.~J.~E.}\ \bibnamefont {Marsh}},\
  }\href {https://arxiv.org/abs/2105.01406} {\enquote {\bibinfo {title} {Axion
  dark matter: What is it and why now?}}\ } (\bibinfo {year} {2022}),\ \Eprint
  {http://arxiv.org/abs/2105.01406} {arXiv:2105.01406 [hep-ph]} \BibitemShut
  {NoStop}%
\bibitem [{\citenamefont {Reyn\'es}\ \emph {et~al.}(2021)\citenamefont
  {Reyn\'es}, \citenamefont {Matthews}, \citenamefont {Reynolds}, \citenamefont
  {Russell}, \citenamefont {Smith},\ and\ \citenamefont
  {Marsh}}]{Reynes:2021bpe}%
  \BibitemOpen
  \bibfield  {author} {\bibinfo {author} {\bibfnamefont {J.~S.}\ \bibnamefont
  {Reyn\'es}}, \bibinfo {author} {\bibfnamefont {J.~H.}\ \bibnamefont
  {Matthews}}, \bibinfo {author} {\bibfnamefont {C.~S.}\ \bibnamefont
  {Reynolds}}, \bibinfo {author} {\bibfnamefont {H.~R.}\ \bibnamefont
  {Russell}}, \bibinfo {author} {\bibfnamefont {R.~N.}\ \bibnamefont {Smith}},
  \ and\ \bibinfo {author} {\bibfnamefont {M.~C.~D.}\ \bibnamefont {Marsh}},\
  }\href {\doibase 10.1093/mnras/stab3464} {\bibfield  {journal} {\bibinfo
  {journal} {Mon. Not. Roy. Astron. Soc.}\ }\textbf {\bibinfo {volume} {510}},\
  \bibinfo {pages} {1264} (\bibinfo {year} {2021})},\ \Eprint
  {http://arxiv.org/abs/2109.03261} {arXiv:2109.03261 [astro-ph.HE]}
  \BibitemShut {NoStop}%
\bibitem [{\citenamefont {Marsh}\ \emph {et~al.}(2017)\citenamefont {Marsh},
  \citenamefont {Russell}, \citenamefont {Fabian}, \citenamefont {McNamara},
  \citenamefont {Nulsen},\ and\ \citenamefont {Reynolds}}]{Marsh_2017}%
  \BibitemOpen
  \bibfield  {author} {\bibinfo {author} {\bibfnamefont {M.~D.}\ \bibnamefont
  {Marsh}}, \bibinfo {author} {\bibfnamefont {H.~R.}\ \bibnamefont {Russell}},
  \bibinfo {author} {\bibfnamefont {A.~C.}\ \bibnamefont {Fabian}}, \bibinfo
  {author} {\bibfnamefont {B.~R.}\ \bibnamefont {McNamara}}, \bibinfo {author}
  {\bibfnamefont {P.}~\bibnamefont {Nulsen}}, \ and\ \bibinfo {author}
  {\bibfnamefont {C.~S.}\ \bibnamefont {Reynolds}},\ }\href {\doibase
  10.1088/1475-7516/2017/12/036} {\bibfield  {journal} {\bibinfo  {journal}
  {Journal of Cosmology and Astroparticle Physics}\ }\textbf {\bibinfo {volume}
  {2017}},\ \bibinfo {pages} {036} (\bibinfo {year} {2017})}\BibitemShut
  {NoStop}%
\bibitem [{\citenamefont {Giannotti}\ \emph {et~al.}(2017)\citenamefont
  {Giannotti}, \citenamefont {Irastorza}, \citenamefont {Redondo},
  \citenamefont {Ringwald},\ and\ \citenamefont {Saikawa}}]{Giannotti:2017hny}%
  \BibitemOpen
  \bibfield  {author} {\bibinfo {author} {\bibfnamefont {M.}~\bibnamefont
  {Giannotti}}, \bibinfo {author} {\bibfnamefont {I.~G.}\ \bibnamefont
  {Irastorza}}, \bibinfo {author} {\bibfnamefont {J.}~\bibnamefont {Redondo}},
  \bibinfo {author} {\bibfnamefont {A.}~\bibnamefont {Ringwald}}, \ and\
  \bibinfo {author} {\bibfnamefont {K.}~\bibnamefont {Saikawa}},\ }\href
  {\doibase 10.1088/1475-7516/2017/10/010} {\bibfield  {journal} {\bibinfo
  {journal} {JCAP}\ }\textbf {\bibinfo {volume} {10}},\ \bibinfo {pages} {010}
  (\bibinfo {year} {2017})},\ \Eprint {http://arxiv.org/abs/1708.02111}
  {arXiv:1708.02111 [hep-ph]} \BibitemShut {NoStop}%
\bibitem [{\citenamefont {Corsico}\ \emph {et~al.}(2012)\citenamefont
  {Corsico}, \citenamefont {Althaus}, \citenamefont {Romero}, \citenamefont
  {Mukadam}, \citenamefont {Garcia-Berro}, \citenamefont {Isern}, \citenamefont
  {Kepler},\ and\ \citenamefont {Corti}}]{Corsico:2012sh}%
  \BibitemOpen
  \bibfield  {author} {\bibinfo {author} {\bibfnamefont {A.~H.}\ \bibnamefont
  {Corsico}}, \bibinfo {author} {\bibfnamefont {L.~G.}\ \bibnamefont
  {Althaus}}, \bibinfo {author} {\bibfnamefont {A.~D.}\ \bibnamefont {Romero}},
  \bibinfo {author} {\bibfnamefont {A.~S.}\ \bibnamefont {Mukadam}}, \bibinfo
  {author} {\bibfnamefont {E.}~\bibnamefont {Garcia-Berro}}, \bibinfo {author}
  {\bibfnamefont {J.}~\bibnamefont {Isern}}, \bibinfo {author} {\bibfnamefont
  {S.~O.}\ \bibnamefont {Kepler}}, \ and\ \bibinfo {author} {\bibfnamefont
  {M.~A.}\ \bibnamefont {Corti}},\ }\href {\doibase
  10.1088/1475-7516/2012/12/010} {\bibfield  {journal} {\bibinfo  {journal}
  {JCAP}\ }\textbf {\bibinfo {volume} {12}},\ \bibinfo {pages} {010} (\bibinfo
  {year} {2012})},\ \Eprint {http://arxiv.org/abs/1211.3389} {arXiv:1211.3389
  [astro-ph.SR]} \BibitemShut {NoStop}%
\bibitem [{\citenamefont {Anastassopoulos}\ \emph {et~al.}(2017)\citenamefont
  {Anastassopoulos} \emph {et~al.}}]{CAST:2017uph}%
  \BibitemOpen
  \bibfield  {author} {\bibinfo {author} {\bibfnamefont {V.}~\bibnamefont
  {Anastassopoulos}} \emph {et~al.} (\bibinfo {collaboration} {CAST}),\ }\href
  {\doibase 10.1038/nphys4109} {\bibfield  {journal} {\bibinfo  {journal}
  {Nature Phys.}\ }\textbf {\bibinfo {volume} {13}},\ \bibinfo {pages} {584}
  (\bibinfo {year} {2017})},\ \Eprint {http://arxiv.org/abs/1705.02290}
  {arXiv:1705.02290 [hep-ex]} \BibitemShut {NoStop}%
\bibitem [{\citenamefont {Ballou}\ \emph {et~al.}(2015)\citenamefont {Ballou},
  \citenamefont {Deferne}, \citenamefont {Finger}, \citenamefont {Finger},
  \citenamefont {Flekova}, \citenamefont {Hosek}, \citenamefont {Kunc},
  \citenamefont {Macuchova}, \citenamefont {Meissner}, \citenamefont {Pugnat},
  \citenamefont {Schott}, \citenamefont {Siemko}, \citenamefont {Slunecka},
  \citenamefont {Sulc}, \citenamefont {Weinsheimer},\ and\ \citenamefont
  {Zicha}}]{PhysRevD.92.092002}%
  \BibitemOpen
  \bibfield  {author} {\bibinfo {author} {\bibfnamefont {R.}~\bibnamefont
  {Ballou}}, \bibinfo {author} {\bibfnamefont {G.}~\bibnamefont {Deferne}},
  \bibinfo {author} {\bibfnamefont {M.}~\bibnamefont {Finger}}, \bibinfo
  {author} {\bibfnamefont {M.}~\bibnamefont {Finger}}, \bibinfo {author}
  {\bibfnamefont {L.}~\bibnamefont {Flekova}}, \bibinfo {author} {\bibfnamefont
  {J.}~\bibnamefont {Hosek}}, \bibinfo {author} {\bibfnamefont
  {S.}~\bibnamefont {Kunc}}, \bibinfo {author} {\bibfnamefont {K.}~\bibnamefont
  {Macuchova}}, \bibinfo {author} {\bibfnamefont {K.~A.}\ \bibnamefont
  {Meissner}}, \bibinfo {author} {\bibfnamefont {P.}~\bibnamefont {Pugnat}},
  \bibinfo {author} {\bibfnamefont {M.}~\bibnamefont {Schott}}, \bibinfo
  {author} {\bibfnamefont {A.}~\bibnamefont {Siemko}}, \bibinfo {author}
  {\bibfnamefont {M.}~\bibnamefont {Slunecka}}, \bibinfo {author}
  {\bibfnamefont {M.}~\bibnamefont {Sulc}}, \bibinfo {author} {\bibfnamefont
  {C.}~\bibnamefont {Weinsheimer}}, \ and\ \bibinfo {author} {\bibfnamefont
  {J.}~\bibnamefont {Zicha}} (\bibinfo {collaboration} {OSQAR Collaboration}),\
  }\href {\doibase 10.1103/PhysRevD.92.092002} {\bibfield  {journal} {\bibinfo
  {journal} {Phys. Rev. D}\ }\textbf {\bibinfo {volume} {92}},\ \bibinfo
  {pages} {092002} (\bibinfo {year} {2015})}\BibitemShut {NoStop}%
\bibitem [{\citenamefont {Melissinos}(2009)}]{PhysRevLett.102.202001}%
  \BibitemOpen
  \bibfield  {author} {\bibinfo {author} {\bibfnamefont {A.~C.}\ \bibnamefont
  {Melissinos}},\ }\href {\doibase 10.1103/PhysRevLett.102.202001} {\bibfield
  {journal} {\bibinfo  {journal} {Phys. Rev. Lett.}\ }\textbf {\bibinfo
  {volume} {102}},\ \bibinfo {pages} {202001} (\bibinfo {year}
  {2009})}\BibitemShut {NoStop}%
\bibitem [{\citenamefont {Liu}\ \emph {et~al.}(2019)\citenamefont {Liu},
  \citenamefont {Elwood}, \citenamefont {Evans},\ and\ \citenamefont
  {Thaler}}]{PhysRevD.100.023548}%
  \BibitemOpen
  \bibfield  {author} {\bibinfo {author} {\bibfnamefont {H.}~\bibnamefont
  {Liu}}, \bibinfo {author} {\bibfnamefont {B.~D.}\ \bibnamefont {Elwood}},
  \bibinfo {author} {\bibfnamefont {M.}~\bibnamefont {Evans}}, \ and\ \bibinfo
  {author} {\bibfnamefont {J.}~\bibnamefont {Thaler}},\ }\href {\doibase
  10.1103/PhysRevD.100.023548} {\bibfield  {journal} {\bibinfo  {journal}
  {Phys. Rev. D}\ }\textbf {\bibinfo {volume} {100}},\ \bibinfo {pages}
  {023548} (\bibinfo {year} {2019})}\BibitemShut {NoStop}%
\bibitem [{\citenamefont {Heinze}\ \emph
  {et~al.}(2024{\natexlab{a}})\citenamefont {Heinze}, \citenamefont {Gill},
  \citenamefont {Dmitriev}, \citenamefont {Smetana}, \citenamefont {Yan},
  \citenamefont {Boyer}, \citenamefont {Martynov},\ and\ \citenamefont
  {Evans}}]{PhysRevLett.132.191002}%
  \BibitemOpen
  \bibfield  {author} {\bibinfo {author} {\bibfnamefont {J.}~\bibnamefont
  {Heinze}}, \bibinfo {author} {\bibfnamefont {A.}~\bibnamefont {Gill}},
  \bibinfo {author} {\bibfnamefont {A.}~\bibnamefont {Dmitriev}}, \bibinfo
  {author} {\bibfnamefont {J.~c.~v.}\ \bibnamefont {Smetana}}, \bibinfo
  {author} {\bibfnamefont {T.}~\bibnamefont {Yan}}, \bibinfo {author}
  {\bibfnamefont {V.}~\bibnamefont {Boyer}}, \bibinfo {author} {\bibfnamefont
  {D.}~\bibnamefont {Martynov}}, \ and\ \bibinfo {author} {\bibfnamefont
  {M.}~\bibnamefont {Evans}},\ }\href {\doibase 10.1103/PhysRevLett.132.191002}
  {\bibfield  {journal} {\bibinfo  {journal} {Phys. Rev. Lett.}\ }\textbf
  {\bibinfo {volume} {132}},\ \bibinfo {pages} {191002} (\bibinfo {year}
  {2024}{\natexlab{a}})}\BibitemShut {NoStop}%
\bibitem [{\citenamefont {Pandey}\ \emph {et~al.}(2024)\citenamefont {Pandey},
  \citenamefont {Hall},\ and\ \citenamefont
  {Evans}}]{pandey2024resultsaxiondarkmatterbirefringent}%
  \BibitemOpen
  \bibfield  {author} {\bibinfo {author} {\bibfnamefont {S.}~\bibnamefont
  {Pandey}}, \bibinfo {author} {\bibfnamefont {E.~D.}\ \bibnamefont {Hall}}, \
  and\ \bibinfo {author} {\bibfnamefont {M.}~\bibnamefont {Evans}},\ }\href
  {https://arxiv.org/abs/2404.12517} {\enquote {\bibinfo {title} {First results
  from the axion dark-matter birefringent cavity (adbc) experiment},}\ }
  (\bibinfo {year} {2024}),\ \Eprint {http://arxiv.org/abs/2404.12517}
  {arXiv:2404.12517 [hep-ex]} \BibitemShut {NoStop}%
\bibitem [{\citenamefont {Martynov}\ and\ \citenamefont
  {Miao}(2020)}]{PhysRevD.101.095034}%
  \BibitemOpen
  \bibfield  {author} {\bibinfo {author} {\bibfnamefont {D.}~\bibnamefont
  {Martynov}}\ and\ \bibinfo {author} {\bibfnamefont {H.}~\bibnamefont
  {Miao}},\ }\href {\doibase 10.1103/PhysRevD.101.095034} {\bibfield  {journal}
  {\bibinfo  {journal} {Phys. Rev. D}\ }\textbf {\bibinfo {volume} {101}},\
  \bibinfo {pages} {095034} (\bibinfo {year} {2020})}\BibitemShut {NoStop}%
\bibitem [{\citenamefont {Ivanov}\ \emph {et~al.}(2019)\citenamefont {Ivanov},
  \citenamefont {Kovalev}, \citenamefont {Lister}, \citenamefont {Panin},
  \citenamefont {Pushkarev}, \citenamefont {Savolainen},\ and\ \citenamefont
  {Troitsky}}]{Ivanov:2018byi}%
  \BibitemOpen
  \bibfield  {author} {\bibinfo {author} {\bibfnamefont {M.~M.}\ \bibnamefont
  {Ivanov}}, \bibinfo {author} {\bibfnamefont {Y.~Y.}\ \bibnamefont {Kovalev}},
  \bibinfo {author} {\bibfnamefont {M.~L.}\ \bibnamefont {Lister}}, \bibinfo
  {author} {\bibfnamefont {A.~G.}\ \bibnamefont {Panin}}, \bibinfo {author}
  {\bibfnamefont {A.~B.}\ \bibnamefont {Pushkarev}}, \bibinfo {author}
  {\bibfnamefont {T.}~\bibnamefont {Savolainen}}, \ and\ \bibinfo {author}
  {\bibfnamefont {S.~V.}\ \bibnamefont {Troitsky}},\ }\href {\doibase
  10.1088/1475-7516/2019/02/059} {\bibfield  {journal} {\bibinfo  {journal}
  {JCAP}\ }\textbf {\bibinfo {volume} {02}},\ \bibinfo {pages} {059} (\bibinfo
  {year} {2019})},\ \Eprint {http://arxiv.org/abs/1811.10997} {arXiv:1811.10997
  [astro-ph.CO]} \BibitemShut {NoStop}%
\bibitem [{\citenamefont {Liu}\ \emph {et~al.}(2020)\citenamefont {Liu},
  \citenamefont {Smoot},\ and\ \citenamefont {Zhao}}]{PhysRevD.101.063012}%
  \BibitemOpen
  \bibfield  {author} {\bibinfo {author} {\bibfnamefont {T.}~\bibnamefont
  {Liu}}, \bibinfo {author} {\bibfnamefont {G.}~\bibnamefont {Smoot}}, \ and\
  \bibinfo {author} {\bibfnamefont {Y.}~\bibnamefont {Zhao}},\ }\href {\doibase
  10.1103/PhysRevD.101.063012} {\bibfield  {journal} {\bibinfo  {journal}
  {Phys. Rev. D}\ }\textbf {\bibinfo {volume} {101}},\ \bibinfo {pages}
  {063012} (\bibinfo {year} {2020})}\BibitemShut {NoStop}%
\bibitem [{\citenamefont {Fedderke}\ \emph {et~al.}(2019)\citenamefont
  {Fedderke}, \citenamefont {Graham},\ and\ \citenamefont
  {Rajendran}}]{PhysRevD.100.015040}%
  \BibitemOpen
  \bibfield  {author} {\bibinfo {author} {\bibfnamefont {M.~A.}\ \bibnamefont
  {Fedderke}}, \bibinfo {author} {\bibfnamefont {P.~W.}\ \bibnamefont
  {Graham}}, \ and\ \bibinfo {author} {\bibfnamefont {S.}~\bibnamefont
  {Rajendran}},\ }\href {\doibase 10.1103/PhysRevD.100.015040} {\bibfield
  {journal} {\bibinfo  {journal} {Phys. Rev. D}\ }\textbf {\bibinfo {volume}
  {100}},\ \bibinfo {pages} {015040} (\bibinfo {year} {2019})}\BibitemShut
  {NoStop}%
\bibitem [{\citenamefont {Plascencia}\ and\ \citenamefont
  {Urbano}(2018)}]{Plascencia:2017kca}%
  \BibitemOpen
  \bibfield  {author} {\bibinfo {author} {\bibfnamefont {A.~D.}\ \bibnamefont
  {Plascencia}}\ and\ \bibinfo {author} {\bibfnamefont {A.}~\bibnamefont
  {Urbano}},\ }\href {\doibase 10.1088/1475-7516/2018/04/059} {\bibfield
  {journal} {\bibinfo  {journal} {JCAP}\ }\textbf {\bibinfo {volume} {04}},\
  \bibinfo {pages} {059} (\bibinfo {year} {2018})},\ \Eprint
  {http://arxiv.org/abs/1711.08298} {arXiv:1711.08298 [gr-qc]} \BibitemShut
  {NoStop}%
\bibitem [{\citenamefont {Fujita}\ \emph {et~al.}(2019)\citenamefont {Fujita},
  \citenamefont {Tazaki},\ and\ \citenamefont {Toma}}]{PhysRevLett.122.191101}%
  \BibitemOpen
  \bibfield  {author} {\bibinfo {author} {\bibfnamefont {T.}~\bibnamefont
  {Fujita}}, \bibinfo {author} {\bibfnamefont {R.}~\bibnamefont {Tazaki}}, \
  and\ \bibinfo {author} {\bibfnamefont {K.}~\bibnamefont {Toma}},\ }\href
  {\doibase 10.1103/PhysRevLett.122.191101} {\bibfield  {journal} {\bibinfo
  {journal} {Phys. Rev. Lett.}\ }\textbf {\bibinfo {volume} {122}},\ \bibinfo
  {pages} {191101} (\bibinfo {year} {2019})}\BibitemShut {NoStop}%
\bibitem [{\citenamefont {Yuan}\ \emph {et~al.}(2021)\citenamefont {Yuan},
  \citenamefont {Xia}, \citenamefont {Tang}, \citenamefont {Zhao},
  \citenamefont {Cai}, \citenamefont {Chen}, \citenamefont {Shu},\ and\
  \citenamefont {Yuan}}]{Yuan:2020xui}%
  \BibitemOpen
  \bibfield  {author} {\bibinfo {author} {\bibfnamefont {G.-W.}\ \bibnamefont
  {Yuan}}, \bibinfo {author} {\bibfnamefont {Z.-Q.}\ \bibnamefont {Xia}},
  \bibinfo {author} {\bibfnamefont {C.}~\bibnamefont {Tang}}, \bibinfo {author}
  {\bibfnamefont {Y.}~\bibnamefont {Zhao}}, \bibinfo {author} {\bibfnamefont
  {Y.-F.}\ \bibnamefont {Cai}}, \bibinfo {author} {\bibfnamefont
  {Y.}~\bibnamefont {Chen}}, \bibinfo {author} {\bibfnamefont {J.}~\bibnamefont
  {Shu}}, \ and\ \bibinfo {author} {\bibfnamefont {Q.}~\bibnamefont {Yuan}},\
  }\href {\doibase 10.1088/1475-7516/2021/03/018} {\bibfield  {journal}
  {\bibinfo  {journal} {JCAP}\ }\textbf {\bibinfo {volume} {03}},\ \bibinfo
  {pages} {018} (\bibinfo {year} {2021})},\ \Eprint
  {http://arxiv.org/abs/2008.13662} {arXiv:2008.13662 [astro-ph.HE]}
  \BibitemShut {NoStop}%
\bibitem [{\citenamefont {Liu}\ \emph {et~al.}(2023)\citenamefont {Liu},
  \citenamefont {Lou},\ and\ \citenamefont {Ren}}]{PhysRevLett.130.121401}%
  \BibitemOpen
  \bibfield  {author} {\bibinfo {author} {\bibfnamefont {T.}~\bibnamefont
  {Liu}}, \bibinfo {author} {\bibfnamefont {X.}~\bibnamefont {Lou}}, \ and\
  \bibinfo {author} {\bibfnamefont {J.}~\bibnamefont {Ren}},\ }\href {\doibase
  10.1103/PhysRevLett.130.121401} {\bibfield  {journal} {\bibinfo  {journal}
  {Phys. Rev. Lett.}\ }\textbf {\bibinfo {volume} {130}},\ \bibinfo {pages}
  {121401} (\bibinfo {year} {2023})}\BibitemShut {NoStop}%
\bibitem [{\citenamefont {DeRocco}\ and\ \citenamefont
  {Hook}(2018)}]{PhysRevD.98.035021}%
  \BibitemOpen
  \bibfield  {author} {\bibinfo {author} {\bibfnamefont {W.}~\bibnamefont
  {DeRocco}}\ and\ \bibinfo {author} {\bibfnamefont {A.}~\bibnamefont {Hook}},\
  }\href {\doibase 10.1103/PhysRevD.98.035021} {\bibfield  {journal} {\bibinfo
  {journal} {Phys. Rev. D}\ }\textbf {\bibinfo {volume} {98}},\ \bibinfo
  {pages} {035021} (\bibinfo {year} {2018})}\BibitemShut {NoStop}%
\bibitem [{\citenamefont {Nagano}\ \emph {et~al.}(2019)\citenamefont {Nagano},
  \citenamefont {Fujita}, \citenamefont {Michimura},\ and\ \citenamefont
  {Obata}}]{PhysRevLett.123.111301}%
  \BibitemOpen
  \bibfield  {author} {\bibinfo {author} {\bibfnamefont {K.}~\bibnamefont
  {Nagano}}, \bibinfo {author} {\bibfnamefont {T.}~\bibnamefont {Fujita}},
  \bibinfo {author} {\bibfnamefont {Y.}~\bibnamefont {Michimura}}, \ and\
  \bibinfo {author} {\bibfnamefont {I.}~\bibnamefont {Obata}},\ }\href
  {\doibase 10.1103/PhysRevLett.123.111301} {\bibfield  {journal} {\bibinfo
  {journal} {Phys. Rev. Lett.}\ }\textbf {\bibinfo {volume} {123}},\ \bibinfo
  {pages} {111301} (\bibinfo {year} {2019})}\BibitemShut {NoStop}%
\bibitem [{\citenamefont {Heinze}\ \emph
  {et~al.}(2024{\natexlab{b}})\citenamefont {Heinze}, \citenamefont {Gill},
  \citenamefont {Dmitriev}, \citenamefont {Smetana}, \citenamefont {Yan},
  \citenamefont {Boyer}, \citenamefont {Martynov}, \citenamefont {Grote},
  \citenamefont {Lough}, \citenamefont {Ejlli},\ and\ \citenamefont
  {Müller}}]{Heinze_2024}%
  \BibitemOpen
  \bibfield  {author} {\bibinfo {author} {\bibfnamefont {J.}~\bibnamefont
  {Heinze}}, \bibinfo {author} {\bibfnamefont {A.}~\bibnamefont {Gill}},
  \bibinfo {author} {\bibfnamefont {A.}~\bibnamefont {Dmitriev}}, \bibinfo
  {author} {\bibfnamefont {J.}~\bibnamefont {Smetana}}, \bibinfo {author}
  {\bibfnamefont {T.}~\bibnamefont {Yan}}, \bibinfo {author} {\bibfnamefont
  {V.}~\bibnamefont {Boyer}}, \bibinfo {author} {\bibfnamefont
  {D.}~\bibnamefont {Martynov}}, \bibinfo {author} {\bibfnamefont
  {H.}~\bibnamefont {Grote}}, \bibinfo {author} {\bibfnamefont
  {J.}~\bibnamefont {Lough}}, \bibinfo {author} {\bibfnamefont
  {A.}~\bibnamefont {Ejlli}}, \ and\ \bibinfo {author} {\bibfnamefont
  {G.}~\bibnamefont {Müller}},\ }\href {\doibase 10.1088/1367-2630/ad48ac}
  {\bibfield  {journal} {\bibinfo  {journal} {New Journal of Physics}\ }\textbf
  {\bibinfo {volume} {26}},\ \bibinfo {pages} {055002} (\bibinfo {year}
  {2024}{\natexlab{b}})}\BibitemShut {NoStop}%
\bibitem [{\citenamefont {Aoki}\ and\ \citenamefont
  {Soda}(2016)}]{Aoki:2016kwl}%
  \BibitemOpen
  \bibfield  {author} {\bibinfo {author} {\bibfnamefont {A.}~\bibnamefont
  {Aoki}}\ and\ \bibinfo {author} {\bibfnamefont {J.}~\bibnamefont {Soda}},\
  }\href {\doibase 10.1142/S0218271817500638} {\bibfield  {journal} {\bibinfo
  {journal} {Int. J. Mod. Phys. D}\ }\textbf {\bibinfo {volume} {26}},\
  \bibinfo {pages} {1750063} (\bibinfo {year} {2016})},\ \Eprint
  {http://arxiv.org/abs/1608.05933} {arXiv:1608.05933 [astro-ph.CO]}
  \BibitemShut {NoStop}%
\bibitem [{\citenamefont {Vermeulen}\ \emph {et~al.}(2021)\citenamefont
  {Vermeulen} \emph {et~al.}}]{Vermeulen:2021epa}%
  \BibitemOpen
  \bibfield  {author} {\bibinfo {author} {\bibfnamefont {S.~M.}\ \bibnamefont
  {Vermeulen}} \emph {et~al.},\ }\href {\doibase 10.1038/s41586-021-04031-y} {\
   (\bibinfo {year} {2021}),\ 10.1038/s41586-021-04031-y},\ \Eprint
  {http://arxiv.org/abs/2103.03783} {arXiv:2103.03783 [gr-qc]} \BibitemShut
  {NoStop}%
\bibitem [{\citenamefont {Pierce}\ \emph {et~al.}(2018)\citenamefont {Pierce},
  \citenamefont {Riles},\ and\ \citenamefont {Zhao}}]{PhysRevLett.121.061102}%
  \BibitemOpen
  \bibfield  {author} {\bibinfo {author} {\bibfnamefont {A.}~\bibnamefont
  {Pierce}}, \bibinfo {author} {\bibfnamefont {K.}~\bibnamefont {Riles}}, \
  and\ \bibinfo {author} {\bibfnamefont {Y.}~\bibnamefont {Zhao}},\ }\href
  {\doibase 10.1103/PhysRevLett.121.061102} {\bibfield  {journal} {\bibinfo
  {journal} {Phys. Rev. Lett.}\ }\textbf {\bibinfo {volume} {121}},\ \bibinfo
  {pages} {061102} (\bibinfo {year} {2018})}\BibitemShut {NoStop}%
\bibitem [{\citenamefont {Abbott}\ \emph {et~al.}(2022)\citenamefont {Abbott}
  \emph {et~al.}}]{LIGOScientific:2021ffg}%
  \BibitemOpen
  \bibfield  {author} {\bibinfo {author} {\bibfnamefont {R.}~\bibnamefont
  {Abbott}} \emph {et~al.} (\bibinfo {collaboration} {LIGO Scientific, KAGRA,
  Virgo}),\ }\href {\doibase 10.1103/PhysRevD.105.063030} {\bibfield  {journal}
  {\bibinfo  {journal} {Phys. Rev. D}\ }\textbf {\bibinfo {volume} {105}},\
  \bibinfo {pages} {063030} (\bibinfo {year} {2022})},\ \bibinfo {note}
  {[Erratum: Phys.Rev.D 109, 089902 (2024)]},\ \Eprint
  {http://arxiv.org/abs/2105.13085} {arXiv:2105.13085 [astro-ph.CO]}
  \BibitemShut {NoStop}%
\bibitem [{\citenamefont {Guo}\ \emph {et~al.}(2019)\citenamefont {Guo},
  \citenamefont {Riles}, \citenamefont {Yang},\ and\ \citenamefont
  {Zhao}}]{Guo:2019ker}%
  \BibitemOpen
  \bibfield  {author} {\bibinfo {author} {\bibfnamefont {H.-K.}\ \bibnamefont
  {Guo}}, \bibinfo {author} {\bibfnamefont {K.}~\bibnamefont {Riles}}, \bibinfo
  {author} {\bibfnamefont {F.-W.}\ \bibnamefont {Yang}}, \ and\ \bibinfo
  {author} {\bibfnamefont {Y.}~\bibnamefont {Zhao}},\ }\href {\doibase
  10.1038/s42005-019-0255-0} {\bibfield  {journal} {\bibinfo  {journal}
  {Commun. Phys.}\ }\textbf {\bibinfo {volume} {2}},\ \bibinfo {pages} {155}
  (\bibinfo {year} {2019})},\ \Eprint {http://arxiv.org/abs/1905.04316}
  {arXiv:1905.04316 [hep-ph]} \BibitemShut {NoStop}%
\bibitem [{\citenamefont {Miller}\ and\ \citenamefont
  {Mendes}(2023)}]{PhysRevD.107.063015}%
  \BibitemOpen
  \bibfield  {author} {\bibinfo {author} {\bibfnamefont {A.~L.}\ \bibnamefont
  {Miller}}\ and\ \bibinfo {author} {\bibfnamefont {L.}~\bibnamefont
  {Mendes}},\ }\href {\doibase 10.1103/PhysRevD.107.063015} {\bibfield
  {journal} {\bibinfo  {journal} {Phys. Rev. D}\ }\textbf {\bibinfo {volume}
  {107}},\ \bibinfo {pages} {063015} (\bibinfo {year} {2023})}\BibitemShut
  {NoStop}%
\bibitem [{\citenamefont {Nakatsuka}\ \emph {et~al.}(2023)\citenamefont
  {Nakatsuka}, \citenamefont {Morisaki}, \citenamefont {Fujita}, \citenamefont
  {Kume}, \citenamefont {Michimura}, \citenamefont {Nagano},\ and\
  \citenamefont {Obata}}]{PhysRevD.108.092010}%
  \BibitemOpen
  \bibfield  {author} {\bibinfo {author} {\bibfnamefont {H.}~\bibnamefont
  {Nakatsuka}}, \bibinfo {author} {\bibfnamefont {S.}~\bibnamefont {Morisaki}},
  \bibinfo {author} {\bibfnamefont {T.}~\bibnamefont {Fujita}}, \bibinfo
  {author} {\bibfnamefont {J.}~\bibnamefont {Kume}}, \bibinfo {author}
  {\bibfnamefont {Y.}~\bibnamefont {Michimura}}, \bibinfo {author}
  {\bibfnamefont {K.}~\bibnamefont {Nagano}}, \ and\ \bibinfo {author}
  {\bibfnamefont {I.}~\bibnamefont {Obata}},\ }\href {\doibase
  10.1103/PhysRevD.108.092010} {\bibfield  {journal} {\bibinfo  {journal}
  {Phys. Rev. D}\ }\textbf {\bibinfo {volume} {108}},\ \bibinfo {pages}
  {092010} (\bibinfo {year} {2023})}\BibitemShut {NoStop}%
\bibitem [{\citenamefont {Fukusumi}\ \emph {et~al.}(2023)\citenamefont
  {Fukusumi}, \citenamefont {Morisaki},\ and\ \citenamefont
  {Suyama}}]{PhysRevD.108.095054}%
  \BibitemOpen
  \bibfield  {author} {\bibinfo {author} {\bibfnamefont {K.}~\bibnamefont
  {Fukusumi}}, \bibinfo {author} {\bibfnamefont {S.}~\bibnamefont {Morisaki}},
  \ and\ \bibinfo {author} {\bibfnamefont {T.}~\bibnamefont {Suyama}},\ }\href
  {\doibase 10.1103/PhysRevD.108.095054} {\bibfield  {journal} {\bibinfo
  {journal} {Phys. Rev. D}\ }\textbf {\bibinfo {volume} {108}},\ \bibinfo
  {pages} {095054} (\bibinfo {year} {2023})}\BibitemShut {NoStop}%
\bibitem [{\citenamefont {Yu}\ \emph {et~al.}(2023)\citenamefont {Yu},
  \citenamefont {Yao}, \citenamefont {Tang},\ and\ \citenamefont
  {Wu}}]{PhysRevD.108.083007}%
  \BibitemOpen
  \bibfield  {author} {\bibinfo {author} {\bibfnamefont {J.-C.}\ \bibnamefont
  {Yu}}, \bibinfo {author} {\bibfnamefont {Y.-H.}\ \bibnamefont {Yao}},
  \bibinfo {author} {\bibfnamefont {Y.}~\bibnamefont {Tang}}, \ and\ \bibinfo
  {author} {\bibfnamefont {Y.-L.}\ \bibnamefont {Wu}},\ }\href {\doibase
  10.1103/PhysRevD.108.083007} {\bibfield  {journal} {\bibinfo  {journal}
  {Phys. Rev. D}\ }\textbf {\bibinfo {volume} {108}},\ \bibinfo {pages}
  {083007} (\bibinfo {year} {2023})}\BibitemShut {NoStop}%
\bibitem [{\citenamefont {Kim}(2023)}]{Kim:2023pkx}%
  \BibitemOpen
  \bibfield  {author} {\bibinfo {author} {\bibfnamefont {H.}~\bibnamefont
  {Kim}},\ }\href {\doibase 10.1088/1475-7516/2023/12/018} {\bibfield
  {journal} {\bibinfo  {journal} {JCAP}\ }\textbf {\bibinfo {volume} {12}},\
  \bibinfo {pages} {018} (\bibinfo {year} {2023})},\ \Eprint
  {http://arxiv.org/abs/2306.13348} {arXiv:2306.13348 [hep-ph]} \BibitemShut
  {NoStop}%
\bibitem [{\citenamefont {Yu}\ \emph {et~al.}(2024)\citenamefont {Yu},
  \citenamefont {Cao}, \citenamefont {Tang},\ and\ \citenamefont
  {Wu}}]{Yu:2024enm}%
  \BibitemOpen
  \bibfield  {author} {\bibinfo {author} {\bibfnamefont {J.-C.}\ \bibnamefont
  {Yu}}, \bibinfo {author} {\bibfnamefont {Y.}~\bibnamefont {Cao}}, \bibinfo
  {author} {\bibfnamefont {Y.}~\bibnamefont {Tang}}, \ and\ \bibinfo {author}
  {\bibfnamefont {Y.-L.}\ \bibnamefont {Wu}},\ }\href {\doibase
  10.1103/PhysRevD.110.023025} {\bibfield  {journal} {\bibinfo  {journal}
  {Phys. Rev. D}\ }\textbf {\bibinfo {volume} {110}},\ \bibinfo {pages}
  {023025} (\bibinfo {year} {2024})},\ \Eprint
  {http://arxiv.org/abs/2404.04333} {arXiv:2404.04333 [hep-ph]} \BibitemShut
  {NoStop}%
\bibitem [{\citenamefont {Yao}\ and\ \citenamefont
  {Tang}(2024)}]{PhysRevD.110.095015}%
  \BibitemOpen
  \bibfield  {author} {\bibinfo {author} {\bibfnamefont {Y.-H.}\ \bibnamefont
  {Yao}}\ and\ \bibinfo {author} {\bibfnamefont {Y.}~\bibnamefont {Tang}},\
  }\href {\doibase 10.1103/PhysRevD.110.095015} {\bibfield  {journal} {\bibinfo
   {journal} {Phys. Rev. D}\ }\textbf {\bibinfo {volume} {110}},\ \bibinfo
  {pages} {095015} (\bibinfo {year} {2024})}\BibitemShut {NoStop}%
\bibitem [{\citenamefont
  {et~al}(2017)}]{amaroseoane2017laserinterferometerspaceantenna}%
  \BibitemOpen
  \bibfield  {author} {\bibinfo {author} {\bibfnamefont {P.~A.-S.}\
  \bibnamefont {et~al}},\ }\href {https://arxiv.org/abs/1702.00786} {\enquote
  {\bibinfo {title} {Laser interferometer space antenna},}\ } (\bibinfo {year}
  {2017}),\ \Eprint {http://arxiv.org/abs/1702.00786} {arXiv:1702.00786
  [astro-ph.IM]} \BibitemShut {NoStop}%
\bibitem [{\citenamefont {Hu}\ and\ \citenamefont {Wu}(2017)}]{Hu:2017mde}%
  \BibitemOpen
  \bibfield  {author} {\bibinfo {author} {\bibfnamefont {W.-R.}\ \bibnamefont
  {Hu}}\ and\ \bibinfo {author} {\bibfnamefont {Y.-L.}\ \bibnamefont {Wu}},\
  }\href {\doibase 10.1093/nsr/nwx116} {\bibfield  {journal} {\bibinfo
  {journal} {Natl. Sci. Rev.}\ }\textbf {\bibinfo {volume} {4}},\ \bibinfo
  {pages} {685} (\bibinfo {year} {2017})}\BibitemShut {NoStop}%
\bibitem [{\citenamefont {et~al}(2016)}]{Luo_2016}%
  \BibitemOpen
  \bibfield  {author} {\bibinfo {author} {\bibfnamefont {J.~L.}\ \bibnamefont
  {et~al}},\ }\href {\doibase 10.1088/0264-9381/33/3/035010} {\bibfield
  {journal} {\bibinfo  {journal} {Classical and Quantum Gravity}\ }\textbf
  {\bibinfo {volume} {33}},\ \bibinfo {pages} {035010} (\bibinfo {year}
  {2016})}\BibitemShut {NoStop}%
\bibitem [{\citenamefont {Crowder}\ and\ \citenamefont
  {Cornish}(2005)}]{Crowder:2005nr}%
  \BibitemOpen
  \bibfield  {author} {\bibinfo {author} {\bibfnamefont {J.}~\bibnamefont
  {Crowder}}\ and\ \bibinfo {author} {\bibfnamefont {N.~J.}\ \bibnamefont
  {Cornish}},\ }\href {\doibase 10.1103/PhysRevD.72.083005} {\bibfield
  {journal} {\bibinfo  {journal} {Phys. Rev. D}\ }\textbf {\bibinfo {volume}
  {72}},\ \bibinfo {pages} {083005} (\bibinfo {year} {2005})},\ \Eprint
  {http://arxiv.org/abs/gr-qc/0506015} {arXiv:gr-qc/0506015} \BibitemShut
  {NoStop}%
\bibitem [{\citenamefont {Carroll}\ \emph {et~al.}(1990)\citenamefont
  {Carroll}, \citenamefont {Field},\ and\ \citenamefont
  {Jackiw}}]{PhysRevD.41.1231}%
  \BibitemOpen
  \bibfield  {author} {\bibinfo {author} {\bibfnamefont {S.~M.}\ \bibnamefont
  {Carroll}}, \bibinfo {author} {\bibfnamefont {G.~B.}\ \bibnamefont {Field}},
  \ and\ \bibinfo {author} {\bibfnamefont {R.}~\bibnamefont {Jackiw}},\ }\href
  {\doibase 10.1103/PhysRevD.41.1231} {\bibfield  {journal} {\bibinfo
  {journal} {Phys. Rev. D}\ }\textbf {\bibinfo {volume} {41}},\ \bibinfo
  {pages} {1231} (\bibinfo {year} {1990})}\BibitemShut {NoStop}%
\bibitem [{\citenamefont {Carroll}\ and\ \citenamefont
  {Field}(1991)}]{PhysRevD.43.3789}%
  \BibitemOpen
  \bibfield  {author} {\bibinfo {author} {\bibfnamefont {S.~M.}\ \bibnamefont
  {Carroll}}\ and\ \bibinfo {author} {\bibfnamefont {G.~B.}\ \bibnamefont
  {Field}},\ }\href {\doibase 10.1103/PhysRevD.43.3789} {\bibfield  {journal}
  {\bibinfo  {journal} {Phys. Rev. D}\ }\textbf {\bibinfo {volume} {43}},\
  \bibinfo {pages} {3789} (\bibinfo {year} {1991})}\BibitemShut {NoStop}%
\bibitem [{\citenamefont {Harari}\ and\ \citenamefont
  {Sikivie}(1992)}]{Harari:1992ea}%
  \BibitemOpen
  \bibfield  {author} {\bibinfo {author} {\bibfnamefont {D.}~\bibnamefont
  {Harari}}\ and\ \bibinfo {author} {\bibfnamefont {P.}~\bibnamefont
  {Sikivie}},\ }\href {\doibase 10.1016/0370-2693(92)91363-E} {\bibfield
  {journal} {\bibinfo  {journal} {Phys. Lett. B}\ }\textbf {\bibinfo {volume}
  {289}},\ \bibinfo {pages} {67} (\bibinfo {year} {1992})}\BibitemShut
  {NoStop}%
\bibitem [{\citenamefont {Khmelnitsky}\ and\ \citenamefont
  {Rubakov}(2014)}]{Khmelnitsky_2014}%
  \BibitemOpen
  \bibfield  {author} {\bibinfo {author} {\bibfnamefont {A.}~\bibnamefont
  {Khmelnitsky}}\ and\ \bibinfo {author} {\bibfnamefont {V.}~\bibnamefont
  {Rubakov}},\ }\href {\doibase 10.1088/1475-7516/2014/02/019} {\bibfield
  {journal} {\bibinfo  {journal} {Journal of Cosmology and Astroparticle
  Physics}\ }\textbf {\bibinfo {volume} {2014}},\ \bibinfo {pages} {019}
  (\bibinfo {year} {2014})}\BibitemShut {NoStop}%
\bibitem [{\citenamefont {Hui}(2021)}]{Hui:2021tkt}%
  \BibitemOpen
  \bibfield  {author} {\bibinfo {author} {\bibfnamefont {L.}~\bibnamefont
  {Hui}},\ }\href {\doibase 10.1146/annurev-astro-120920-010024} {\bibfield
  {journal} {\bibinfo  {journal} {Ann. Rev. Astron. Astrophys.}\ }\textbf
  {\bibinfo {volume} {59}},\ \bibinfo {pages} {247} (\bibinfo {year} {2021})},\
  \Eprint {http://arxiv.org/abs/2101.11735} {arXiv:2101.11735 [astro-ph.CO]}
  \BibitemShut {NoStop}%
\bibitem [{\citenamefont {Tinto}\ and\ \citenamefont
  {Dhurandhar}(2021)}]{Tinto:2020fcc}%
  \BibitemOpen
  \bibfield  {author} {\bibinfo {author} {\bibfnamefont {M.}~\bibnamefont
  {Tinto}}\ and\ \bibinfo {author} {\bibfnamefont {S.~V.}\ \bibnamefont
  {Dhurandhar}},\ }\href {\doibase 10.1007/s41114-020-00029-6} {\bibfield
  {journal} {\bibinfo  {journal} {Living Rev. Rel.}\ }\textbf {\bibinfo
  {volume} {24}},\ \bibinfo {pages} {1} (\bibinfo {year} {2021})}\BibitemShut
  {NoStop}%
\bibitem [{\citenamefont {Armstrong}\ \emph {et~al.}(1999)\citenamefont
  {Armstrong}, \citenamefont {Estabrook},\ and\ \citenamefont
  {Tinto}}]{Armstrong_1999}%
  \BibitemOpen
  \bibfield  {author} {\bibinfo {author} {\bibfnamefont {J.~W.}\ \bibnamefont
  {Armstrong}}, \bibinfo {author} {\bibfnamefont {F.~B.}\ \bibnamefont
  {Estabrook}}, \ and\ \bibinfo {author} {\bibfnamefont {M.}~\bibnamefont
  {Tinto}},\ }\href {\doibase 10.1086/308110} {\bibfield  {journal} {\bibinfo
  {journal} {The Astrophysical Journal}\ }\textbf {\bibinfo {volume} {527}},\
  \bibinfo {pages} {814} (\bibinfo {year} {1999})}\BibitemShut {NoStop}%
\bibitem [{\citenamefont {Estabrook}\ \emph {et~al.}(2000)\citenamefont
  {Estabrook}, \citenamefont {Tinto},\ and\ \citenamefont
  {Armstrong}}]{PhysRevD.62.042002}%
  \BibitemOpen
  \bibfield  {author} {\bibinfo {author} {\bibfnamefont {F.~B.}\ \bibnamefont
  {Estabrook}}, \bibinfo {author} {\bibfnamefont {M.}~\bibnamefont {Tinto}}, \
  and\ \bibinfo {author} {\bibfnamefont {J.~W.}\ \bibnamefont {Armstrong}},\
  }\href {\doibase 10.1103/PhysRevD.62.042002} {\bibfield  {journal} {\bibinfo
  {journal} {Phys. Rev. D}\ }\textbf {\bibinfo {volume} {62}},\ \bibinfo
  {pages} {042002} (\bibinfo {year} {2000})}\BibitemShut {NoStop}%
\bibitem [{\citenamefont {Dhurandhar}\ \emph {et~al.}(2002)\citenamefont
  {Dhurandhar}, \citenamefont {Nayak},\ and\ \citenamefont
  {Vinet}}]{PhysRevD.65.102002}%
  \BibitemOpen
  \bibfield  {author} {\bibinfo {author} {\bibfnamefont {S.~V.}\ \bibnamefont
  {Dhurandhar}}, \bibinfo {author} {\bibfnamefont {K.~R.}\ \bibnamefont
  {Nayak}}, \ and\ \bibinfo {author} {\bibfnamefont {J.-Y.}\ \bibnamefont
  {Vinet}},\ }\href {\doibase 10.1103/PhysRevD.65.102002} {\bibfield  {journal}
  {\bibinfo  {journal} {Phys. Rev. D}\ }\textbf {\bibinfo {volume} {65}},\
  \bibinfo {pages} {102002} (\bibinfo {year} {2002})}\BibitemShut {NoStop}%
\bibitem [{\citenamefont {Vallisneri}(2005)}]{PhysRevD.72.042003}%
  \BibitemOpen
  \bibfield  {author} {\bibinfo {author} {\bibfnamefont {M.}~\bibnamefont
  {Vallisneri}},\ }\href {\doibase 10.1103/PhysRevD.72.042003} {\bibfield
  {journal} {\bibinfo  {journal} {Phys. Rev. D}\ }\textbf {\bibinfo {volume}
  {72}},\ \bibinfo {pages} {042003} (\bibinfo {year} {2005})}\BibitemShut
  {NoStop}%
\bibitem [{\citenamefont {Foster}\ \emph {et~al.}(2021)\citenamefont {Foster},
  \citenamefont {Kahn}, \citenamefont {Nguyen}, \citenamefont {Rodd},\ and\
  \citenamefont {Safdi}}]{PhysRevD.103.076018}%
  \BibitemOpen
  \bibfield  {author} {\bibinfo {author} {\bibfnamefont {J.~W.}\ \bibnamefont
  {Foster}}, \bibinfo {author} {\bibfnamefont {Y.}~\bibnamefont {Kahn}},
  \bibinfo {author} {\bibfnamefont {R.}~\bibnamefont {Nguyen}}, \bibinfo
  {author} {\bibfnamefont {N.~L.}\ \bibnamefont {Rodd}}, \ and\ \bibinfo
  {author} {\bibfnamefont {B.~R.}\ \bibnamefont {Safdi}},\ }\href {\doibase
  10.1103/PhysRevD.103.076018} {\bibfield  {journal} {\bibinfo  {journal}
  {Phys. Rev. D}\ }\textbf {\bibinfo {volume} {103}},\ \bibinfo {pages}
  {076018} (\bibinfo {year} {2021})}\BibitemShut {NoStop}%
\bibitem [{\citenamefont {Lisanti}\ \emph {et~al.}(2021)\citenamefont
  {Lisanti}, \citenamefont {Moschella},\ and\ \citenamefont
  {Terrano}}]{PhysRevD.104.055037}%
  \BibitemOpen
  \bibfield  {author} {\bibinfo {author} {\bibfnamefont {M.}~\bibnamefont
  {Lisanti}}, \bibinfo {author} {\bibfnamefont {M.}~\bibnamefont {Moschella}},
  \ and\ \bibinfo {author} {\bibfnamefont {W.}~\bibnamefont {Terrano}},\ }\href
  {\doibase 10.1103/PhysRevD.104.055037} {\bibfield  {journal} {\bibinfo
  {journal} {Phys. Rev. D}\ }\textbf {\bibinfo {volume} {104}},\ \bibinfo
  {pages} {055037} (\bibinfo {year} {2021})}\BibitemShut {NoStop}%
\bibitem [{\citenamefont {Gramolin}\ \emph {et~al.}(2022)\citenamefont
  {Gramolin}, \citenamefont {Wickenbrock}, \citenamefont {Aybas}, \citenamefont
  {Bekker}, \citenamefont {Budker}, \citenamefont {Centers}, \citenamefont
  {Figueroa}, \citenamefont {Jackson~Kimball},\ and\ \citenamefont
  {Sushkov}}]{PhysRevD.105.035029}%
  \BibitemOpen
  \bibfield  {author} {\bibinfo {author} {\bibfnamefont {A.~V.}\ \bibnamefont
  {Gramolin}}, \bibinfo {author} {\bibfnamefont {A.}~\bibnamefont
  {Wickenbrock}}, \bibinfo {author} {\bibfnamefont {D.}~\bibnamefont {Aybas}},
  \bibinfo {author} {\bibfnamefont {H.}~\bibnamefont {Bekker}}, \bibinfo
  {author} {\bibfnamefont {D.}~\bibnamefont {Budker}}, \bibinfo {author}
  {\bibfnamefont {G.~P.}\ \bibnamefont {Centers}}, \bibinfo {author}
  {\bibfnamefont {N.~L.}\ \bibnamefont {Figueroa}}, \bibinfo {author}
  {\bibfnamefont {D.~F.}\ \bibnamefont {Jackson~Kimball}}, \ and\ \bibinfo
  {author} {\bibfnamefont {A.~O.}\ \bibnamefont {Sushkov}},\ }\href {\doibase
  10.1103/PhysRevD.105.035029} {\bibfield  {journal} {\bibinfo  {journal}
  {Phys. Rev. D}\ }\textbf {\bibinfo {volume} {105}},\ \bibinfo {pages}
  {035029} (\bibinfo {year} {2022})}\BibitemShut {NoStop}%
\bibitem [{\citenamefont {Amaral}\ \emph {et~al.}(2024)\citenamefont {Amaral},
  \citenamefont {Jain}, \citenamefont {Amin},\ and\ \citenamefont
  {Tunnell}}]{Amaral:2024tjg}%
  \BibitemOpen
  \bibfield  {author} {\bibinfo {author} {\bibfnamefont {D.~W.~P.}\
  \bibnamefont {Amaral}}, \bibinfo {author} {\bibfnamefont {M.}~\bibnamefont
  {Jain}}, \bibinfo {author} {\bibfnamefont {M.~A.}\ \bibnamefont {Amin}}, \
  and\ \bibinfo {author} {\bibfnamefont {C.}~\bibnamefont {Tunnell}},\ }\href
  {\doibase 10.1088/1475-7516/2024/06/050} {\bibfield  {journal} {\bibinfo
  {journal} {JCAP}\ }\textbf {\bibinfo {volume} {06}},\ \bibinfo {pages} {050}
  (\bibinfo {year} {2024})},\ \Eprint {http://arxiv.org/abs/2403.02381}
  {arXiv:2403.02381 [hep-ph]} \BibitemShut {NoStop}%
\bibitem [{\citenamefont {Saleh}\ and\ \citenamefont {Teich}(2019)}]{book}%
  \BibitemOpen
  \bibfield  {author} {\bibinfo {author} {\bibfnamefont {B.}~\bibnamefont
  {Saleh}}\ and\ \bibinfo {author} {\bibfnamefont {M.}~\bibnamefont {Teich}},\
  }\href@noop {} {\emph {\bibinfo {title} {Fundamentals of Photonics, 3rd
  Edition}}}\ (\bibinfo {year} {2019})\BibitemShut {NoStop}%
\bibitem [{\citenamefont {Budker}\ \emph {et~al.}(2014)\citenamefont {Budker},
  \citenamefont {Graham}, \citenamefont {Ledbetter}, \citenamefont
  {Rajendran},\ and\ \citenamefont {Sushkov}}]{PhysRevX.4.021030}%
  \BibitemOpen
  \bibfield  {author} {\bibinfo {author} {\bibfnamefont {D.}~\bibnamefont
  {Budker}}, \bibinfo {author} {\bibfnamefont {P.~W.}\ \bibnamefont {Graham}},
  \bibinfo {author} {\bibfnamefont {M.}~\bibnamefont {Ledbetter}}, \bibinfo
  {author} {\bibfnamefont {S.}~\bibnamefont {Rajendran}}, \ and\ \bibinfo
  {author} {\bibfnamefont {A.~O.}\ \bibnamefont {Sushkov}},\ }\href {\doibase
  10.1103/PhysRevX.4.021030} {\bibfield  {journal} {\bibinfo  {journal} {Phys.
  Rev. X}\ }\textbf {\bibinfo {volume} {4}},\ \bibinfo {pages} {021030}
  (\bibinfo {year} {2014})}\BibitemShut {NoStop}%
\bibitem [{\citenamefont {Babak}\ \emph {et~al.}(2021)\citenamefont {Babak},
  \citenamefont {Hewitson},\ and\ \citenamefont
  {Petiteau}}]{babak2021lisasensitivitysnrcalculations}%
  \BibitemOpen
  \bibfield  {author} {\bibinfo {author} {\bibfnamefont {S.}~\bibnamefont
  {Babak}}, \bibinfo {author} {\bibfnamefont {M.}~\bibnamefont {Hewitson}}, \
  and\ \bibinfo {author} {\bibfnamefont {A.}~\bibnamefont {Petiteau}},\ }\href
  {https://arxiv.org/abs/2108.01167} {\enquote {\bibinfo {title} {Lisa
  sensitivity and snr calculations},}\ } (\bibinfo {year} {2021}),\ \Eprint
  {http://arxiv.org/abs/2108.01167} {arXiv:2108.01167 [astro-ph.IM]}
  \BibitemShut {NoStop}%
\bibitem [{\citenamefont {Corbin}\ and\ \citenamefont
  {Cornish}(2006)}]{Corbin:2005ny}%
  \BibitemOpen
  \bibfield  {author} {\bibinfo {author} {\bibfnamefont {V.}~\bibnamefont
  {Corbin}}\ and\ \bibinfo {author} {\bibfnamefont {N.~J.}\ \bibnamefont
  {Cornish}},\ }\href {\doibase 10.1088/0264-9381/23/7/014} {\bibfield
  {journal} {\bibinfo  {journal} {Class. Quant. Grav.}\ }\textbf {\bibinfo
  {volume} {23}},\ \bibinfo {pages} {2435} (\bibinfo {year} {2006})},\ \Eprint
  {http://arxiv.org/abs/gr-qc/0512039} {arXiv:gr-qc/0512039} \BibitemShut
  {NoStop}%
\bibitem [{\citenamefont {Centers}\ \emph {et~al.}(2021)\citenamefont {Centers}
  \emph {et~al.}}]{Centers:2019dyn}%
  \BibitemOpen
  \bibfield  {author} {\bibinfo {author} {\bibfnamefont {G.~P.}\ \bibnamefont
  {Centers}} \emph {et~al.},\ }\href {\doibase 10.1038/s41467-021-27632-7}
  {\bibfield  {journal} {\bibinfo  {journal} {Nature Commun.}\ }\textbf
  {\bibinfo {volume} {12}},\ \bibinfo {pages} {7321} (\bibinfo {year}
  {2021})},\ \Eprint {http://arxiv.org/abs/1905.13650} {arXiv:1905.13650
  [astro-ph.CO]} \BibitemShut {NoStop}%
\bibitem [{\citenamefont {Payez}\ \emph {et~al.}(2015)\citenamefont {Payez},
  \citenamefont {Evoli}, \citenamefont {Fischer}, \citenamefont {Giannotti},
  \citenamefont {Mirizzi},\ and\ \citenamefont {Ringwald}}]{Payez_2015}%
  \BibitemOpen
  \bibfield  {author} {\bibinfo {author} {\bibfnamefont {A.}~\bibnamefont
  {Payez}}, \bibinfo {author} {\bibfnamefont {C.}~\bibnamefont {Evoli}},
  \bibinfo {author} {\bibfnamefont {T.}~\bibnamefont {Fischer}}, \bibinfo
  {author} {\bibfnamefont {M.}~\bibnamefont {Giannotti}}, \bibinfo {author}
  {\bibfnamefont {A.}~\bibnamefont {Mirizzi}}, \ and\ \bibinfo {author}
  {\bibfnamefont {A.}~\bibnamefont {Ringwald}},\ }\href {\doibase
  10.1088/1475-7516/2015/02/006} {\bibfield  {journal} {\bibinfo  {journal}
  {Journal of Cosmology and Astroparticle Physics}\ }\textbf {\bibinfo {volume}
  {2015}},\ \bibinfo {pages} {006} (\bibinfo {year} {2015})}\BibitemShut
  {NoStop}%
\bibitem [{\citenamefont {et~al}(2006)}]{Kawamura_2006}%
  \BibitemOpen
  \bibfield  {author} {\bibinfo {author} {\bibfnamefont {S.~K.}\ \bibnamefont
  {et~al}},\ }\href {\doibase 10.1088/0264-9381/23/8/S17} {\bibfield  {journal}
  {\bibinfo  {journal} {Classical and Quantum Gravity}\ }\textbf {\bibinfo
  {volume} {23}},\ \bibinfo {pages} {S125} (\bibinfo {year}
  {2006})}\BibitemShut {NoStop}%
\bibitem [{\citenamefont {Martens}\ \emph {et~al.}(2023)\citenamefont
  {Martens}, \citenamefont {Khan},\ and\ \citenamefont {Bayle}}]{Martens_2023}%
  \BibitemOpen
  \bibfield  {author} {\bibinfo {author} {\bibfnamefont {W.}~\bibnamefont
  {Martens}}, \bibinfo {author} {\bibfnamefont {M.}~\bibnamefont {Khan}}, \
  and\ \bibinfo {author} {\bibfnamefont {J.-B.}\ \bibnamefont {Bayle}},\ }\href
  {\doibase 10.1088/1361-6382/acf3c7} {\bibfield  {journal} {\bibinfo
  {journal} {Classical and Quantum Gravity}\ }\textbf {\bibinfo {volume}
  {40}},\ \bibinfo {pages} {195022} (\bibinfo {year} {2023})}\BibitemShut
  {NoStop}%
\bibitem [{\citenamefont {NI}(2013)}]{ASTROD-GW}%
  \BibitemOpen
  \bibfield  {author} {\bibinfo {author} {\bibfnamefont {W.-T.}\ \bibnamefont
  {NI}},\ }\href {\doibase 10.1142/S0218271813410046} {\bibfield  {journal}
  {\bibinfo  {journal} {International Journal of Modern Physics D}\ }\textbf
  {\bibinfo {volume} {22}},\ \bibinfo {pages} {1341004} (\bibinfo {year}
  {2013})},\ \Eprint
  {http://arxiv.org/abs/https://doi.org/10.1142/S0218271813410046}
  {https://doi.org/10.1142/S0218271813410046} \BibitemShut {NoStop}%
\bibitem [{\citenamefont {Harry}\ \emph {et~al.}(2006)\citenamefont {Harry},
  \citenamefont {Fritschel}, \citenamefont {Shaddock}, \citenamefont
  {Folkner},\ and\ \citenamefont {Phinney}}]{Harry:2006fi}%
  \BibitemOpen
  \bibfield  {author} {\bibinfo {author} {\bibfnamefont {G.~M.}\ \bibnamefont
  {Harry}}, \bibinfo {author} {\bibfnamefont {P.}~\bibnamefont {Fritschel}},
  \bibinfo {author} {\bibfnamefont {D.~A.}\ \bibnamefont {Shaddock}}, \bibinfo
  {author} {\bibfnamefont {W.}~\bibnamefont {Folkner}}, \ and\ \bibinfo
  {author} {\bibfnamefont {E.~S.}\ \bibnamefont {Phinney}},\ }\href {\doibase
  10.1088/0264-9381/23/15/008} {\bibfield  {journal} {\bibinfo  {journal}
  {Class. Quant. Grav.}\ }\textbf {\bibinfo {volume} {23}},\ \bibinfo {pages}
  {4887} (\bibinfo {year} {2006})},\ \bibinfo {note} {[Erratum:
  Class.Quant.Grav. 23, 7361 (2006)]}\BibitemShut {NoStop}%
\bibitem [{\citenamefont {Armengaud}\ \emph {et~al.}(2019)\citenamefont
  {Armengaud} \emph {et~al.}}]{IAXO:2019mpb}%
  \BibitemOpen
  \bibfield  {author} {\bibinfo {author} {\bibfnamefont {E.}~\bibnamefont
  {Armengaud}} \emph {et~al.} (\bibinfo {collaboration} {IAXO}),\ }\href
  {\doibase 10.1088/1475-7516/2019/06/047} {\bibfield  {journal} {\bibinfo
  {journal} {JCAP}\ }\textbf {\bibinfo {volume} {06}},\ \bibinfo {pages} {047}
  (\bibinfo {year} {2019})},\ \Eprint {http://arxiv.org/abs/1904.09155}
  {arXiv:1904.09155 [hep-ph]} \BibitemShut {NoStop}%
\bibitem [{\citenamefont {Gue}\ \emph {et~al.}(2024)\citenamefont {Gue},
  \citenamefont {Hees},\ and\ \citenamefont {Wolf}}]{wolf2024}%
  \BibitemOpen
  \bibfield  {author} {\bibinfo {author} {\bibfnamefont {J.}~\bibnamefont
  {Gue}}, \bibinfo {author} {\bibfnamefont {A.}~\bibnamefont {Hees}}, \ and\
  \bibinfo {author} {\bibfnamefont {P.}~\bibnamefont {Wolf}},\ }\href
  {https://arxiv.org/abs/2410.17763} {\enquote {\bibinfo {title} {Probing the
  axion-photon coupling with space-based gravitational waves detectors},}\ }
  (\bibinfo {year} {2024}),\ \Eprint {http://arxiv.org/abs/2410.17763}
  {arXiv:2410.17763 [hep-ph]} \BibitemShut {NoStop}%
\end{thebibliography}%
\end{document}